%% file: final_draft.tex
\documentclass[a4paper,12pt]{article}
\usepackage{graphics}
\usepackage{latexsym}

\font\thirteenrm = cmr12 at 13pt 

\title{ \thirteenrm\textbf{Unequal Intra-layer Coupling in a Bilayer Driven Lattice Gas} }

\author{ 
Choon-Peng Chng and Jian-Sheng Wang}

\date{
\textit{Department of Computational Science, National University of Singapore,} \\
\textit{Singapore 119260, Republic of Singapore.}
\\
\ 
\\
28 October 1999
}
\begin{document}

\maketitle
\begin{abstract}
The system under study is a twin-layered square lattice gas at
half-filling, being driven to non-equilibrium steady states by a
large, finite `electric' field. By making intra-layer couplings
unequal we were able to extend the phase diagram obtained by Hill, Zia
and Schmittmann (1996) and found that the tri-critical point, which
separates the phase regions of the stripped (S) phase (stable at
positive interlayer interactions $J_3$), the filled-empty (FE) phase
(stable at negative $J_3$) and disorder (D), is shifted even further
into the negative $J_3$ region as the coupling traverse to the driving
field increases. Many transient phases to the S phase at the S-FE
boundary were found to be long-lived. We also attempted to test
whether the universality class of D-FE transitions under a drive is
still Ising. Simulation results suggest a value of 1.75 for the
exponent $\gamma$ but a value close to 2.0 for the ratio
$\gamma/\nu$. We speculate that the D-FE second order transition is
different from Ising near criticality, where observed first-order-like
transitions between FE and its ``local minimum" cousin occur during
each simulation run.

\noindent
PACS number(s): 05.50.+q, 64.60.C, 05.70.Jk.
\end{abstract}

\section{INTRODUCTION}

	Equilibrium statistical mechanics has served us well in the
understanding of collective behaviour in many-body systems in, or
near, thermal equilibrium. However, nature abounds with examples of
systems that are far from equilibrium and their behaviour cannot be
predicted by the theory. Linear response theory, a form of
perturbation theory, works well only for systems slightly off
equilibrium but not for those far from equilibrium. The way to tackle
such new systems is to study simple models that have well-understood
equilibrium properties.

	Much work had followed from the early attempt by Katz,
Lebowitz and Spohn \cite{katz:DLG} to drive the Ising lattice gas
model into non-equilibrium steady states via the introduction of an
`external electric field'. This driven lattice gas (DLG) model became
the prototype to study Driven Diffusive Systems (DDS). The
time-independent final state of the DLG model has a probability
distribution which is not given by the usual Boltzmann factor but
depends on the dynamics controlling the evolution.

	The KLS or standard model for a DDS is composed of an ordinary
lattice gas in contact with a thermal bath, having particles hopping
to their nearest-neighbour unoccupied sites. This is controlled by a
rate specified by both the energetics of inter-particle interactions
and an external, uniform driving field \cite{sch:sm_dds}.

	Achahbar and Marro \cite{ach:decoupled} studied a variant of
the standard model: stacking two fully periodic standard models on top
of each other, without interactions across the layers.  This system is
coupled to a heat bath at temperature $T$ using spin-exchange
(Kawasaki) dynamics with the usual Metropolis rate. In Kawasaki
dynamics, pairs of sites (both intra- and inter-layer) are considered
for exchange in order to have a global conservation of particles.
Thus we have a diffusive system without sources or sinks. Half-filled
systems are studied in order to access the critical point. The two
decoupled Ising systems gave two phase transitions as the temperature
is decreased from a large value.  First, the disordered (D) phase at
high $T$ transforms into a state with strips in both layers (S phase).
This is much like two aligned, single-layer driven systems. Upon
further lowering of $T$, a first-order transition occurs which results
in an ordered state, resembling the equilibrium Ising system.  It
consists of a homogeneously filled layer and an empty layer (FE
phase).

	Hill, Zia and Schmittmann \cite{hill:bilayer} unveils the
mystery for the presence of the two phase transitions. They did a
natural extension to Achahbar and Marro's model: addition of a
coupling across the layers. This coupling, $J_z$, can be both
attractive and repulsive. This led to novel discoveries. From the new
phase diagram in $T$--$J_z$ space at a fixed $E$, we can observe the
intrusion of the S phase into that for the FE phase. Please refer to
their paper for the figure.  It was shown that the `usual' FE to D
transition is interrupted by the presence of the S phase.  The two
phase transitions reported by Achahbar {\it et.al.} is located along
the $J_z=0$ line.  Note that the strength of the `electric field' $E$
used is large but finite to drive the system far out of equilibrium.

	In our paper, we investigate such systems further with yet
another trivial modification.  We attempt to observe the effects of
having an unequal coupling in the $x$- and $y$-directions within each
top and bottom layers. In particular, we wish to map out the phase
diagram in the $T$--$J_z$--$J_y$ plane.  Taking {\bf $E$} to be in the
$x$-direction, we have particle-particle interactions in the
transverse direction, $J_y$, being larger or equal to that along the
field, $J_x$.  The latter case should recover Hill {\it et. al}'s
results.

	Besides extending the phase diagram in a new `dimension', we
also attempt to determine the universality class of the system for
$J_z < 0$, i.e. for FE to D second-order transitions.  It was stated
in \cite{hill:bilayer} that preliminary results seem to suggest that
D-S transition belongs to the class of the {\it single-layer driven
lattice gas}. It is our objective here to test the hypothesis that the
D-FE transition belongs to the {\it Ising} universality class, which
many systems belong.

\section{DEFINITION OF THE MODEL AND TOOLS EMPLOYED}

	Following Hill et.al., our system consists of two fully
periodic $L \times L$ square lattices, arranged in a bilayer
structure. We label the sites by $(j_1,j_2,j_3)$ with
$j_1,j_2=0,\ldots,L-1$ and $j_3=0,1$. Each site may be either occupied
or empty, such that we can specify a configuration of the system by a
set of occupation numbers $\{n(j_1,j_2,j_3)\}$, where $n$ is 0 or
1. In spin language, we have spin, $s\ =\ 2n-1\ =\ \pm 1$. For
half-filled systems, $\sum n = L^2$ or $\sum s = 0$ i.e. zero net
magnetization.  The Hamiltonian is given by,
\begin{equation}
H = -J_{1}\sum_{x-dir}nn' - J_{2}\sum_{y-dir}nn'- J_{3}\sum nn'',
\label{hamiltonian}
\end{equation}
where $n$ and $n'$ are the occupancies for nearest
neighbours within a given layer while $n$, $n''$ are for those across
layers. Summations in x- and y-directions include both top and bottom
layers.  Hereupon, $J_{1,2,3}$ will be used in place of $J_{x,y,z}$.

Note that with $J_3=0$, we have two decoupled Ising systems. This has
been confirmed by computing the equivalent Ising model heat capacity
from the system and comparing with exact results, where good agreement
is observed. We restrict $J_1$ and $J_2$ to positive values, with
$J_3/J_1 = \beta$ in the range $[-10,10]$.  For $J_2/J_1 = \alpha$, we
let it take on values 1, 2, 5 and 10. We set $J_1$ to unity and with
$\alpha=1$, we should be able to reproduce results obtained by Hill
et. al.

	The temperature $T$ is given in units of the single layer
Onsager temperature, being $0.5673 J_1/k_B$ in particle
language. Finally, the external driving field $E$ is given in units of
$J_1$ as well, which affects the Metropolis rate via a subtraction of
$E$ from $\triangle H$ for hops along the field and vice-versa. A
value of 25$J_1$ is used throughout the study.

	Lattice sizes investigated are of dimensions $L$= 32, 64 and
128.  Typical Monte Carlo steps (MCS) per site taken are 500,000 for
the phase diagram determination and $10^6$ for the universality class
investigation.  Runs are performed at fixed $Js$, $E$ and $T$,
starting from a random initial configuration generated by a 64-bit
Linear Congruential random number generator. Discarding the first
$5\times 10^4$ MCS, measurements are taken every 200 MCS. We thus
believe that after this amount of steps, the system has settled into a
steady state.  However, if a significant change in character is seen
in the configuration, as in any approach to the true steady state from
any local minimum (in energy), the time average is taken only after
the changeover point.

To determine the critical temperatures, many systems are started from
identical initial states but with different temperature settings. A
susceptibility plot is then constructed from which the $T$ value
giving the maximum susceptibility ($T_{peak}$) is obtained via a
quadratic least-squares fit.  This is to be repeated for each $L$ and
the estimate for $T_c$ obtained via the usual finite-size scaling
hypothesis,
\begin{equation}
T_{peak}(L) - T_c \propto L^{-1/\nu}.
\label{Tc_determ}
\end{equation}

The critical exponent $\nu$ is chosen to be 1.0, as for the Ising
model. In fact, for an undriven system with $J_1=J_2$ and $J_3=0$, the
$T_c$ obtained via this method is 0.9886, using $L$ = 4, 8, 16 and 32.
This is in good agreement with the expected value of 1.0. However, for
a driven system, it has yet to be shown explicitly that $\nu$ is still
1.0, which is the other objective of this paper.

For the D-S transitions, it was suggested in \cite{hill:bilayer} and
\cite{marro:decoupled2} that the critical exponent $\nu$ is
0.7. Nonetheless, due to the enormous demand on computer time,
$T_{peak}$ is taken as a rough estimate for $T_c$ in the determination
of the phase diagrams.  Thus for D-FE, the $T_{peak}$ values serve as
upper bounds on the true critical temperatures. Hence the value of
$\nu$ does not affect the shapes of the phase diagrams significantly.

The susceptibility is defined as,
\begin{equation}
\chi(l_1,l_2,l_3) = {L^d \over {k_B T}}[ \langle |\tilde{n}(l_1,l_2,l_3)|^2 \rangle - 
\langle |\tilde{n}(l_1,l_2,l_3)| \rangle^2], \ k_B = 1,
\label{chi_eqn}
\end{equation}
where $d = 2$ for our 2-D system and
$\langle|\tilde{n}|\rangle$ is taken to be the relevant order
parameter.  We define $\{l_1,l_2,l_3\}$ as taking the same range as
$\{j_1,j_2,j_3\}$ introduced earlier. The Fourier Transform of the
occupancy $n(j_1,j_2,j_3)$ is given by,
\begin{equation}
\tilde{n}(l_{1},l_{2},l_{3}) = 
{1\over{2L^{2}}} {\sum_{j_{1},j_{2},j_{3}} n(j_{1},j_{2},j_{3})e^{2\pi i[(j_{1}l_{1}+j_{2}l_{2})/L + 
j_{3}l_{3}/2]}}.                \label{eq:ft_n}
\end{equation}
Thus in order for the Fast Fourier Transform to be
applicable, only system sizes $L=2^k$ is used, with $k$ being any
positive integer.

The quantity $\langle |\tilde{n}(l_1,l_2,l_3)|^2 \rangle$ is called
the {\bf structure factor}. A change across the lattices is reflected
in the third index, $l_3$, in $\tilde{n}(0,0,1)$. For a perfect FE
phase, $|\tilde{n}(0,0,1)|^2=0.5^2 = 0.25$ is the only non-trivial
positive entry in the power spectrum, besides the trivial
$|\tilde{n}(0,0,0)|^2=0.25$ due to the half-filled nature of the
lattice. Thus the quantity $S(0,0,1)$ computed is the structure factor
for the FE phase, where the time average operations are redundant for
the pure phases.  Other entries in the power spectrum such as
$|\tilde{n}(0,1,0)|^2$ can be used to characterise other phases.  In
fact, Hill {\it et. al.} used this entry's time average S(0,1,0) to
represent the S phase, but we found that any odd $l_2$ index suffices.

\begin{figure}
%the Picture environment:
\begin{picture}(450,200)

% produces a numbered grid, which facilitates the positioning of objects tremendously.
%\graphpaper(0,0)(400,180)
{\small

\put(20,20){\line(0,1){50}}
\put(20,70){\line(1,0){50}}
\put(70,70){\line(0,-1){50}}
\put(70,20){\line(-1,0){50}}
\put(25,45){Empty}

\put(20,90){\line(0,1){50}}
\put(20,140){\line(1,0){50}}
\put(70,140){\line(0,-1){50}}
\put(70,90){\line(-1,0){50}}
\put(25,115){Full}

\put(25,160){FE phase}
\put(10,180){\vector(1,0){25}}
\put(40,180){y-dir}
\put(10,180){\vector(0,-1){25}}
\put(5,145){x-dir}

\put(72,100){$S(0,0,1)$}
\put(72,85){$=(1/2)^2$}

\put(150,20){\line(0,1){50}}
\put(150,70){\line(1,0){50}}
\put(200,70){\line(0,-1){50}}
\put(200,20){\line(-1,0){50}}
\put(175,20){\line(0,1){50}}
%\put(155,45){Full}
%\put(180,45){Empty}
\put(155,25){\shortstack[l]{F\\u\\l\\l}}
\put(180,25){\shortstack[l]{E\\m\\p\\t\\y}}

\put(150,90){\line(0,1){50}}
\put(150,140){\line(1,0){50}}
\put(200,140){\line(0,-1){50}}
\put(200,90){\line(-1,0){50}}
\put(175,90){\line(0,1){50}}
\put(155,95){\shortstack[l]{F\\u\\l\\l}}
\put(180,95){\shortstack[l]{E\\m\\p\\t\\y}}

\put(155,160){S phase}
\put(205,100){$S(0,l_2,0)$}
\put(205,85){$={1 \over {[L\sin{(l_2\pi/L)}]}^2}$}
\put(205,65){$l_2$ is odd integer}

\put(320,20){\line(0,1){50}}
\put(320,70){\line(1,0){50}}
\put(370,70){\line(0,-1){50}}
\put(370,20){\line(-1,0){50}}
\put(345,20){\line(0,1){50}}
\put(325,25){\shortstack[l]{E\\m\\p\\t\\y}}
\put(350,25){\shortstack[l]{F\\u\\l\\l}}

\put(320,90){\line(0,1){50}}
\put(320,140){\line(1,0){50}}
\put(370,140){\line(0,-1){50}}
\put(370,90){\line(-1,0){50}}
\put(345,90){\line(0,1){50}}
\put(325,95){\shortstack[l]{F\\u\\l\\l}}
\put(350,95){\shortstack[l]{E\\m\\p\\t\\y}}

\put(325,160){AFS phase}
\put(375,100){$S(0,l_2,1)$}
\put(375,85){$= S(0,l_2,0)$}
}

\end{picture}
\caption{The `pure' configurations: FE, S and AFS phases.}
\label{fig:pures}
\end{figure}
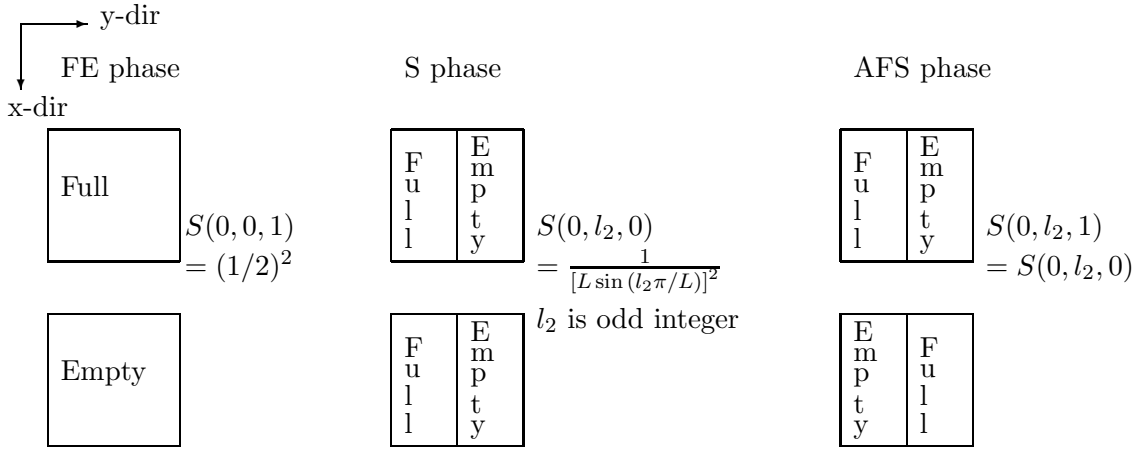

We thus speculate that any given configuration of the bilayer DLG can
be viewed as consisting of a superposition of many `pure tones', such
as the FE configuration. Thus, through a Fourier Transform, we can
pick out the `frequencies' present by monitoring a few entries in the
power spectrum which represent various possible steady states from
energy arguments. Upon taking time averages, the corresponding
structure factors can be computed. For D-FE transitions, S(0,0,1) is
monitored together with S(0,1,1) which represents the `local minimum'
solution. This is a `staggered' form of the FE phase, with an occupied
band on one layer matched by an empty one on the other, which we
termed the AFS(Anti-Ferromagnetic Strip) phase.  It is like a hybrid
between FE and S phases and occurs at low temperatures for systems
with repulsive interlayer coupling. See Fig.\ref{fig:pures} below for
a pictorial view.  The transition to D from a pure FE phase (dominant
at moderate temperatures) is marked by a drop of S(0,0,1) from its
maximum of 0.25 to near zero. The location of $T_c$ is where the slope
of drop is the largest or where $\chi(0,0,1)$ peaks.

Due to finite-size effects, the peaks of the susceptibility function
do not diverge to infinity but is ``rounded" and the peak location
shifted in temperature. These two features are observed from our
simulation data.

\section{NEW PHASE DIAGRAMS}

\begin{figure}
\resizebox{8cm}{!}{\includegraphics{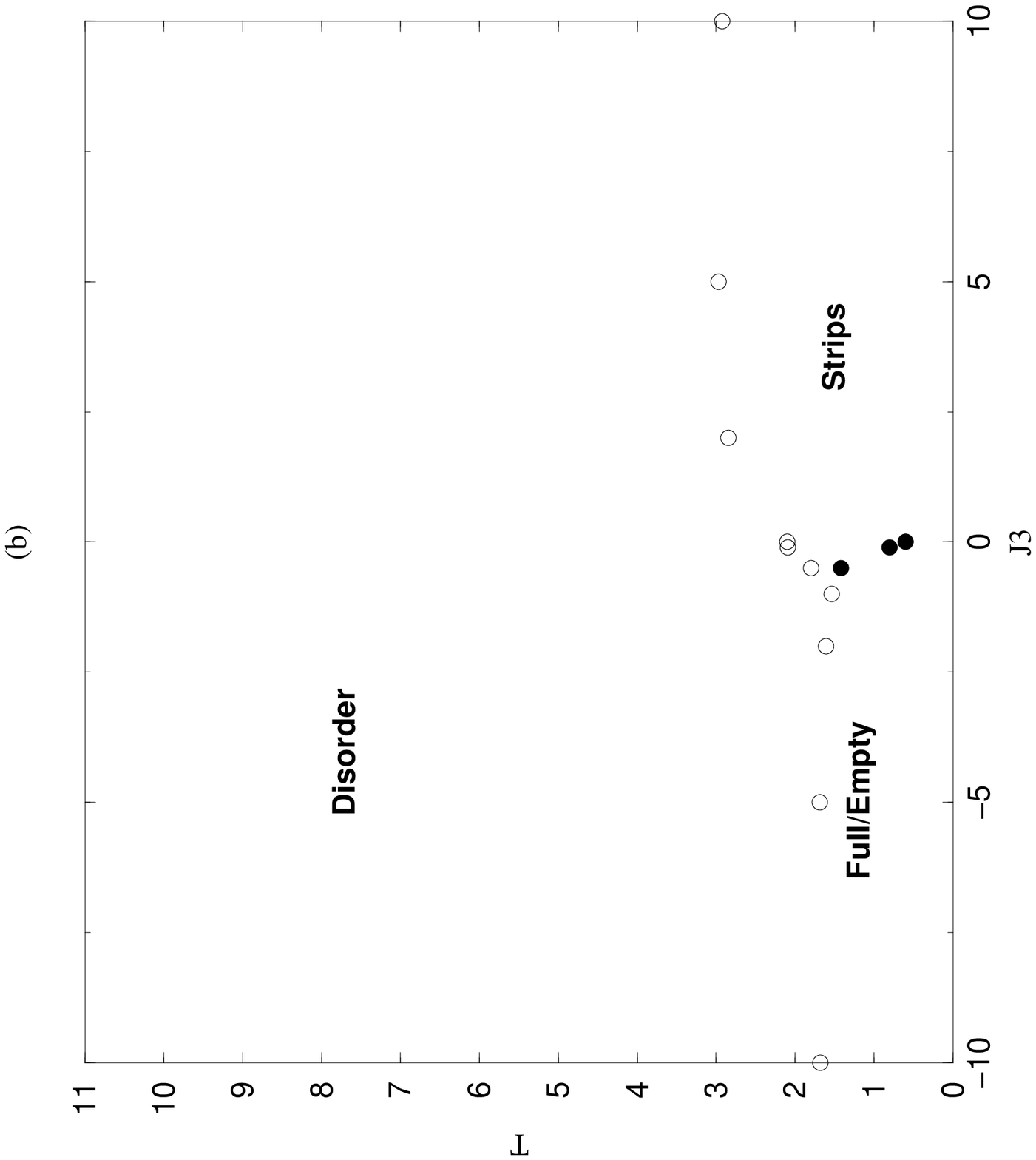}}%
\resizebox{8cm}{!}{\includegraphics{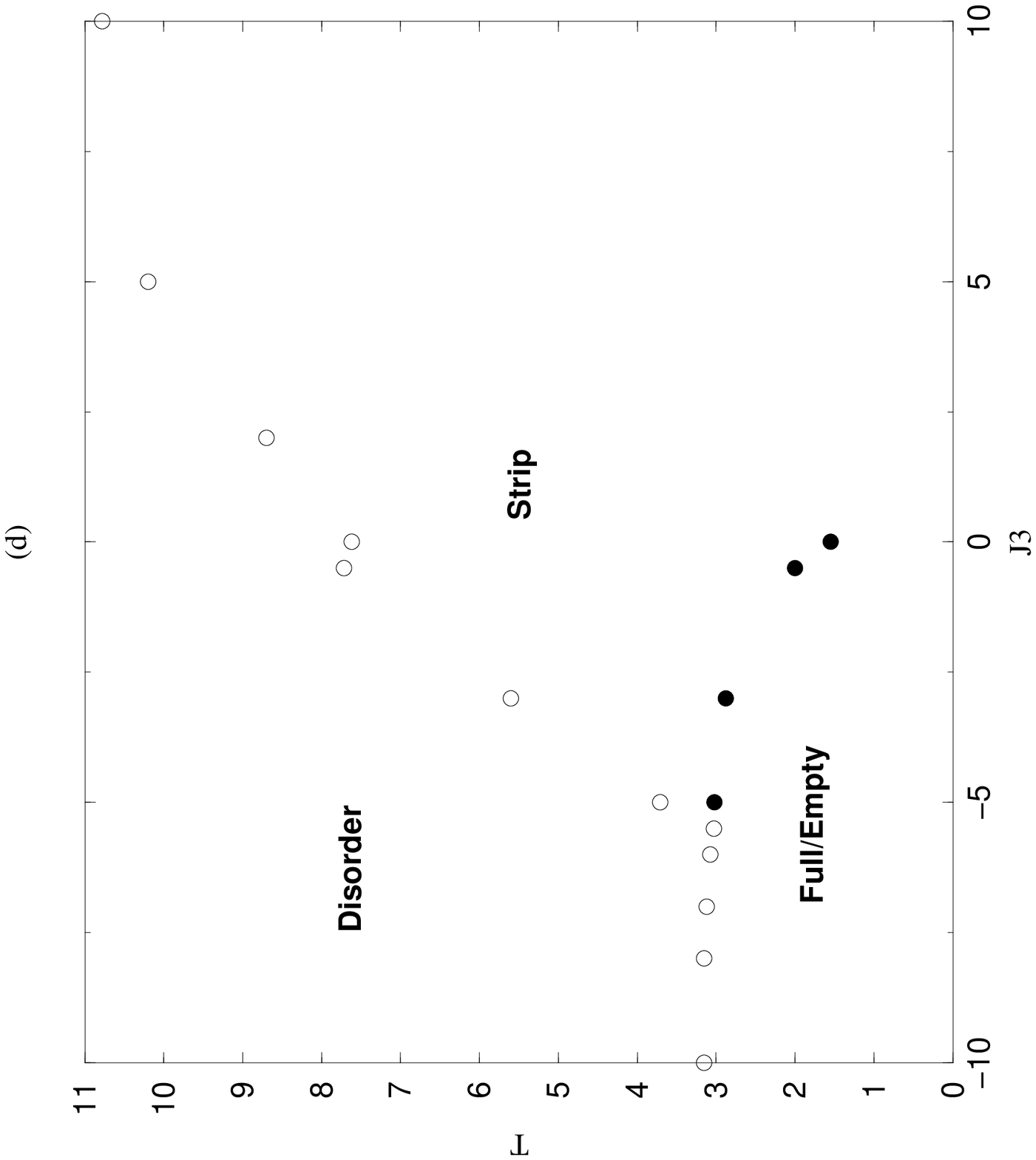}}
\resizebox{8cm}{!}{\includegraphics{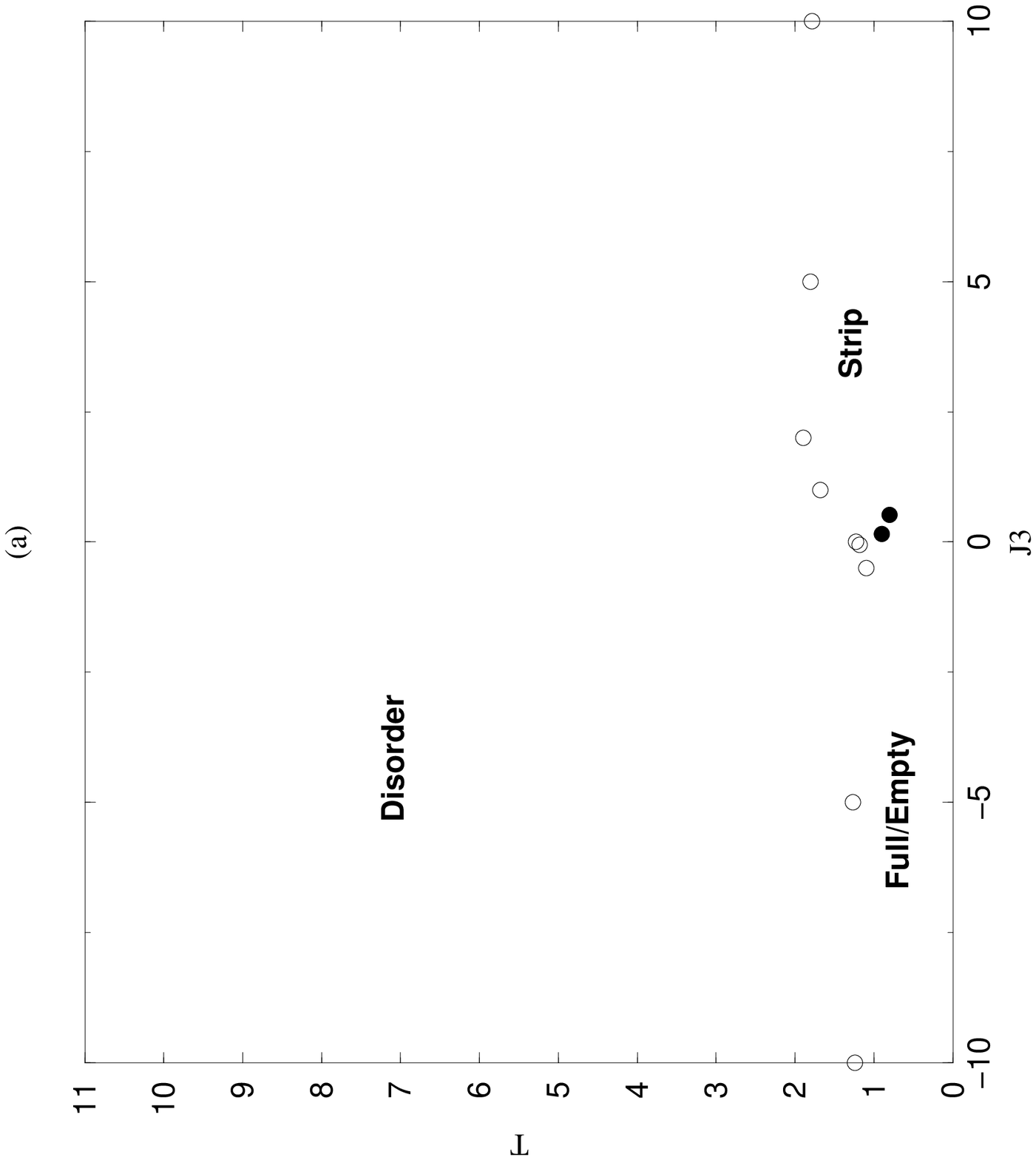}}%
\resizebox{8cm}{!}{\includegraphics{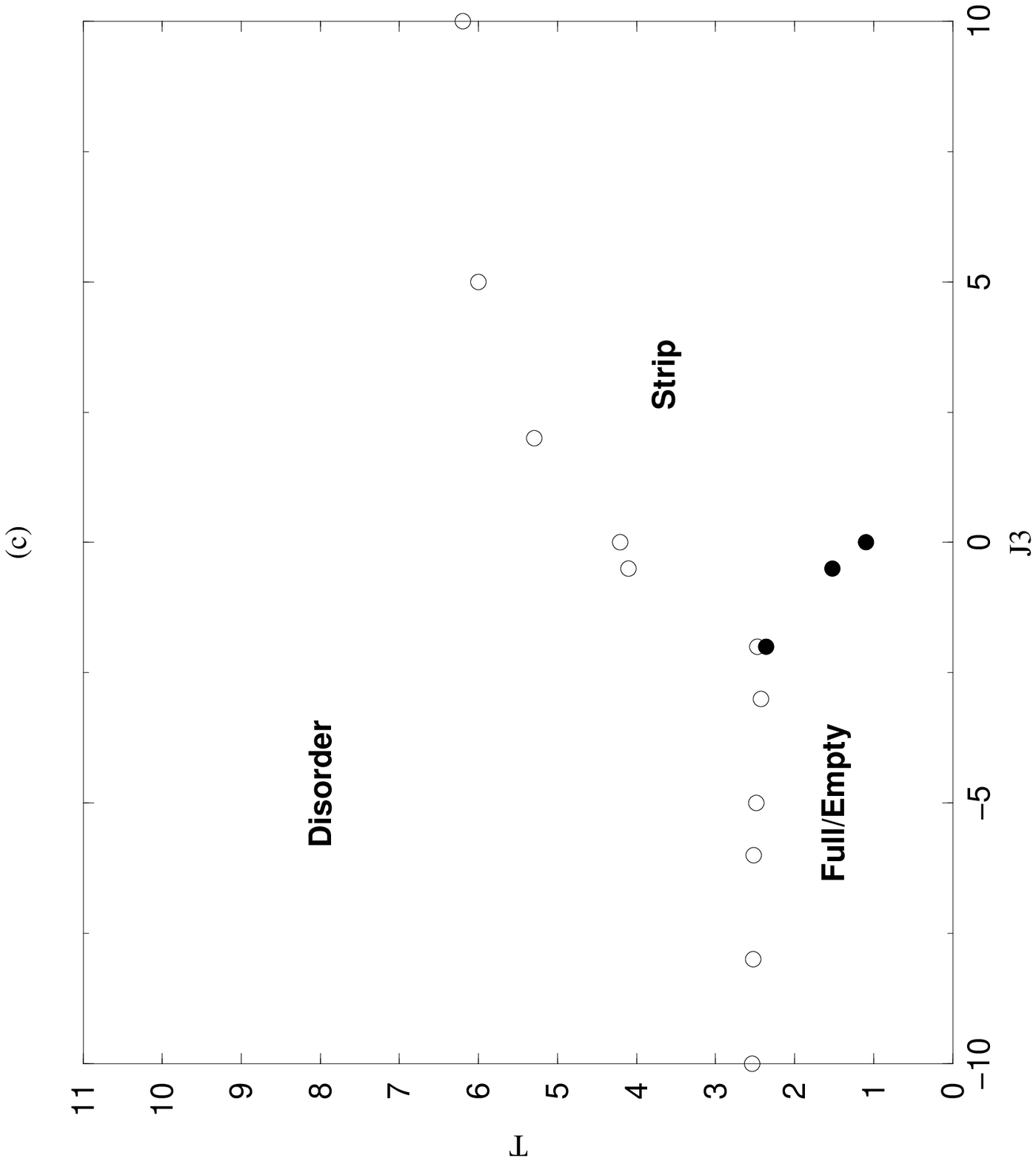}}
\caption{Phase diagrams for the bi-layer lattice gas: $\alpha$ = 1 (a), $\alpha$ = 2 (b),
 $\alpha$ = 5 (c) and  $\alpha$ = 10 (d).} \label{fig:phase_diag}
\end{figure}

The phase diagram for a driven system with the same parameters as used
by Hill et.al. can be reproduced to an acceptable degree by our
implementation, which is of paramount importance to our work here.  We
shall present our finding as a set of four new phase diagrams, including
the one similar to that obtained by Hill's group. The diagrams are
actually slices off the full 3D phase diagram in the $T-\beta-\alpha$
space. Note the $J_3$ will be used interchangeably with $\beta$ for
clearer physical meaning.  See Figure~\ref{fig:phase_diag} for the
phase diagrams, which were all plotted on the same scale for better
comparison.

A few qualitative features can be discerned from the phase
diagrams. The first of these is the growth of the `triangular' region,
a term coined in Hill's paper for the intrusion region of the S phase
into that for the FE phase, as $\alpha$ is increased. This observation
proved beyond doubt that the small `triangular' region seen by Hill is
not an artifact.  Without an external drive, no bias exists between
the FE and the S phases. However, application of a drive in the
x-direction (vertical) seems to favour the S phase with its linear
interface aligned with the drive as compared to the isotropic FE
phase. This is speculated to be analogous to magnetic domain growth in
a ferromagnetic material under the action of an external magnetic
field. The S phase, which is not expected to be stable when replusive
interactions exist between the layers, could become stable due to the
drive. The driving field could somehow compensate for the gain in
configuration energy as a result of particle stacking under repulsive
interactions.  The survival of the S phase in the negative $\beta$
$(=J_3/J_1)$ region is increased as the coupling transverse to the
drive ($J_2$) increases.  The phase region occupied by the S phase
thus grows in the expense of the FE phase!

Another feature worth noting is the shifting of the tri-critical point
towards more negative $\beta$ values as well as towards higher
temperatures. Thus the S phase becomes more stable at moderate $\beta$
values as $\alpha$ is increased, despite its instability from energy
arguments.

\begin{figure}
\resizebox{8cm}{!}{\includegraphics{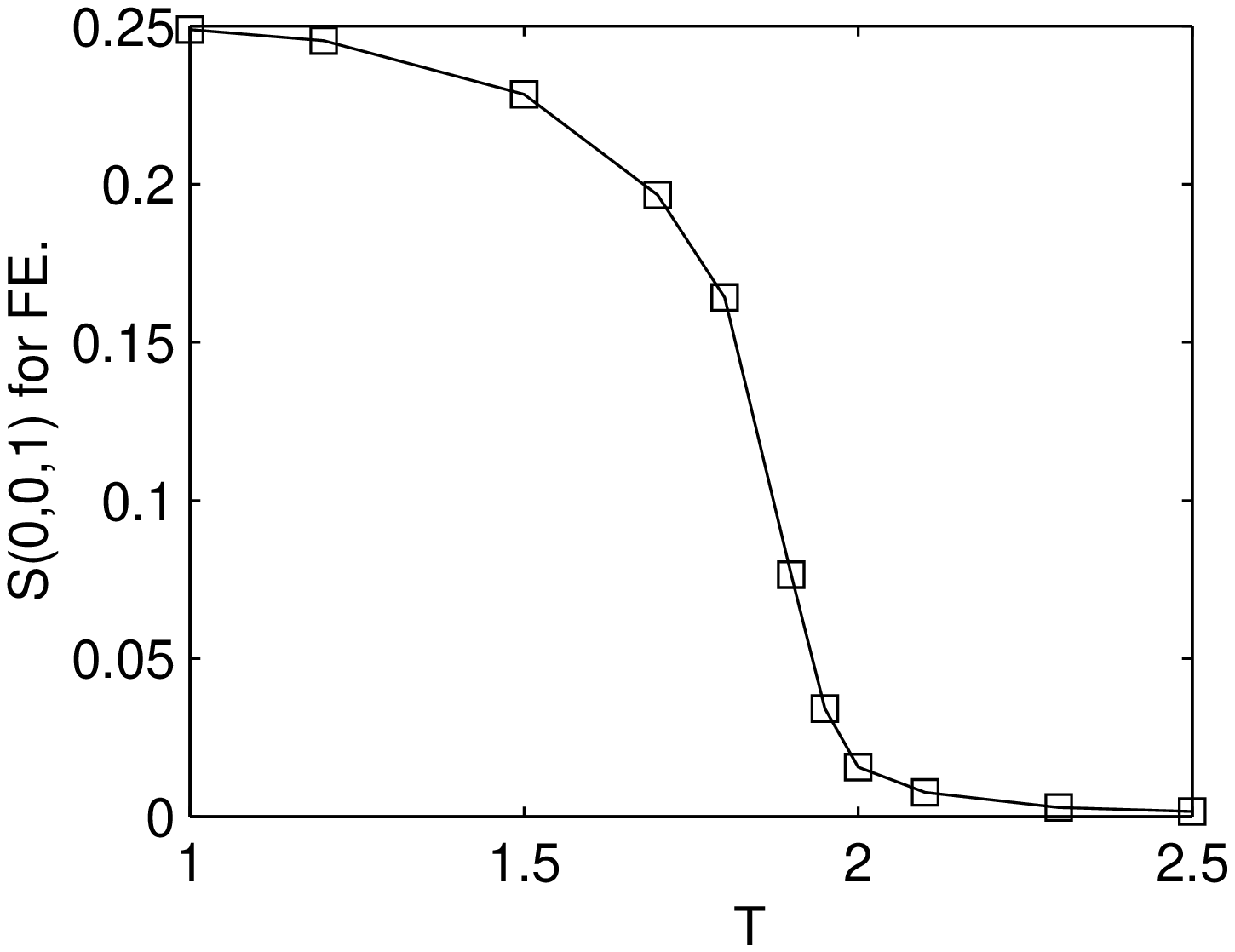}}
\resizebox{8cm}{!}{\includegraphics{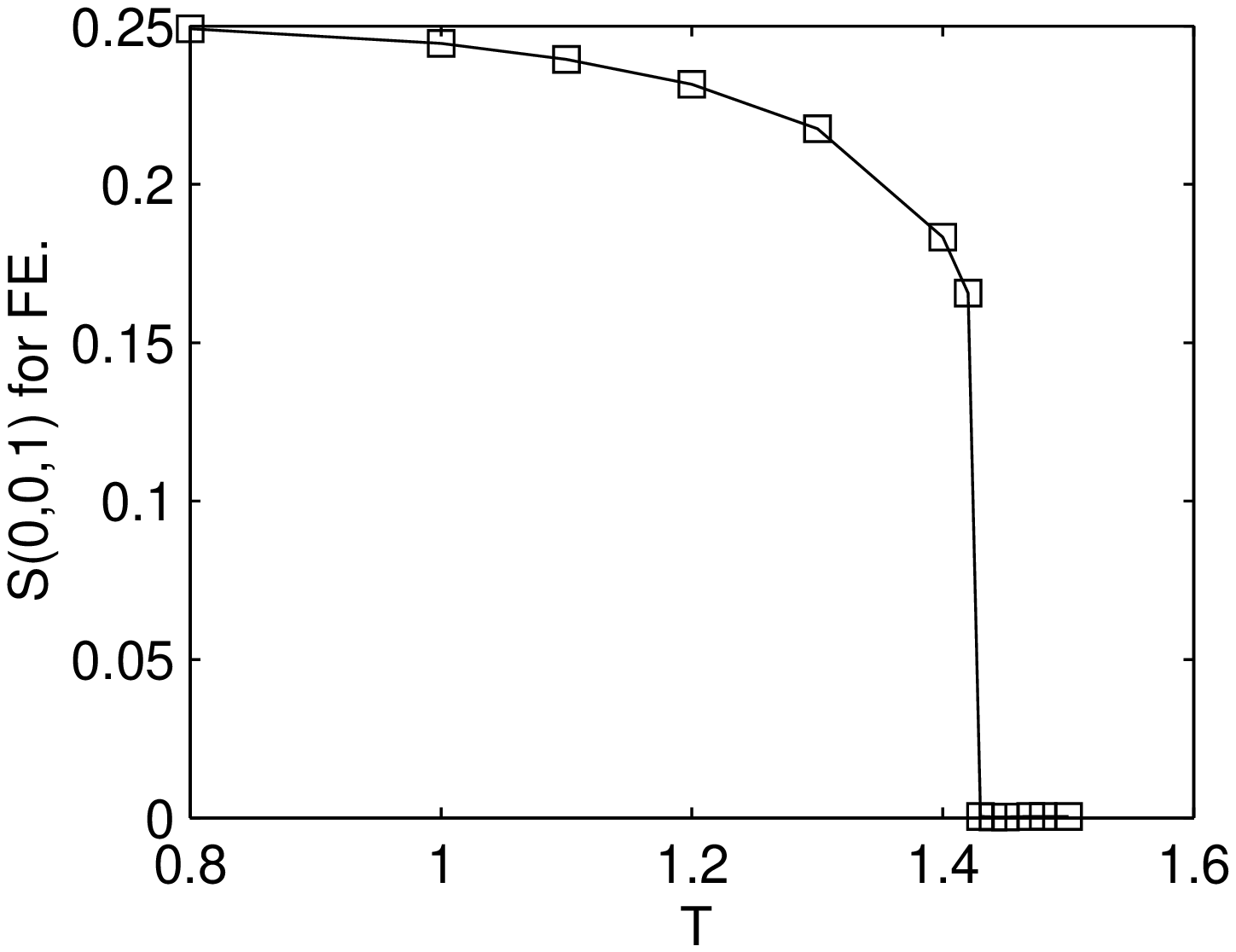}}
\caption{ Plots of structure factor S(0,0,1) (for FE) against $T$.  The
left figure is for $J_{1}=J_{2}=1, J_{3}=-5$, $L=32$, $E=0$ which is
an example of a 2nd order transition.  On the other hand, the right
figure serves as an illustration of the structure factor discontinuity
at a first order transition and is obtained at $J_1 = 1, J_2 = 2, J_3
= -0.5, L = 32$ and $E_x = 25$.  }
\label{fig:graph_strc}
\end{figure}

We judge whether the transition is second or first order by looking at
the plots of structure factors against temperature $T$. A second order
transition has continuous derivatives at every point, an example of
which is shown in Fig. \ref{fig:graph_strc} for the D-FE transition. A
first order transition, like the S-FE, will show a discontinuity as
the right plot in Fig.\ref{fig:graph_strc} illustrates.

Table \ref{table:Tc's} below presents some representative $T_c$ values
from the phase diagrams. One can plot the difference between the $T_c$
values for the 2nd(D-S) [column 4 of table: 0.0(2)] and 1st(D-FE)
[column 3 of table] order transitions along the $\beta$ = 0 line
against $\alpha$ and observe that a least-squares straight line can be
fitted through them. However, due to a lack of finite-sized scaling
knowledge for the D-S transition, we could not get a better estimate
for $T_c$ at the 2nd order transition point and thus could not
conclude if the error bars could tolerate a linear fit. Nonetheless, a
linear fit might be possible, though no theory has yet been developed
to investigate this.

% Have a table of experimental values.
\begin{table}
\begin{tabular}{|r|l|l|l|l|}   \hline
$J_{2}/J_{1}$  &  \multicolumn{4}{c|}{$J_{3}/J_{1}$} \\ \cline{2-5}
		  &  $-$10.0  &  0.0 (1)     &  0.0 (2) &  10.0 \\ \hline\hline
1.0               &  1.2424  & $\sim$0.95 & 1.22828  & 1.7810 \\ \hline
2.0               &  1.6920  & 0.6         & 2.10     & 2.923 \\ \hline
5.0               &  2.5354  & 1.10            & 4.2050  &  6.2  \\ \hline
10.0              &  3.1515  & 1.55            & 7.6190  & 10.779 \\ \hline
\end{tabular}

\caption{Transition temperatures at the three prominent $J_3/J_1$ values.}
\label{table:Tc's}
\end{table}

We also plotted $T_c$ at $\beta = -10$, 0 and 10 against the $\alpha$
values. The plot for $\beta = -10$ (D-FE) seems to exhibit a
logarithmic relationship. As for the non-negative $\beta$ values,
which are for D-S transitions, the relationship seems linear except at
large $\alpha$ for $\beta=10$ and small $\alpha$ for $\beta = 0$.

\section{INTERPRETATIONS AND DISCUSSIONS}
    
The fact that the FE phase survives under a driving field should not
be taken as expected. For large $J_2$ couplings, we would expect
staggered and horizontal anti-ferromagnetic bands to form in the
undriven bilayer DLG from energy arguments. The form looks like the
AFS configuration but rotated by 90 degrees. Under a driving field
directed perpendicular to these bands, it appears that even a large
coupling of 10 could not stand up to the effect of an even larger
driving field (strength 20).  It has yet to be seen if the reverse
situation can favour the rotated AFS phase.

We would like to suggest some explanations for the observations from
the phase diagrams.

Firstly, the increased intrusion of the S phase as $J_2$ increases can
be understood as follows.  In a `thought model', the S phase can be
thought of as consisting of strings of particles, of one particle
width, aligned with the external and large driving field. These are
bounded together through the coupling $J_2$, in the transverse
direction to the field E{\bf $\hat{x}$}. As $T$ increases, the
arrangement will be disturbed till at a sufficiently large $T$,
disorder reigns. However, if we increase $J_2$, the increased binding
could compensate for the disorientating effect of large $T$.  This
effectively makes the critical temperature between S and D phases
higher.

However, this is not to say that the increased $J_2$ does not help to
increase $T_c$ for the D-FE transitions as well. In fact, on careful
observation of the phase diagrams, it does! The increase of $T_c$ was
much smaller in the FE case.

The effect of $J_2$ helps neighbouring particles to bind together in
the $y$-direction. This helps the configuration to hold together
despite the larger temperatures applied and is true for both S and FE
phases. One possible reason for the much lower thermal tolerance for
the FE case might be that each of the $L^2$ particles in one layer has
equal probability to leave the pure FE phase.  On the other hand, for
the S phase, only particles at the edges (top and bottom layers)
aligned with the field can migrate traverse to the field and leave the
pure S phase. This implies that only $2L$ particles has a chance of
migration. Thus it is easier to destroy a FE phase than an S phase
once they are formed. In actual simulations starting from random
configurations, this implies that it is easier to form the S phase.
This might provide the key to the stark difference in the amount of
benefit acquired from an increased $J_2$ for the two pure phases. The
argument also holds for configurations of a `near FE' or `near S'
nature, before the critical temperature.

Further, each movement of a particle out of the filled band for an S
phase violates the occupied-occupied single site configuration across
the layers, which is typical of the S phase. However, the `exchange'
of a particle with a hole on the opposite layer in a FE phase does not
violate the empty-occupied configuration typical of an FE phase! Note
that this argument is only for a single site. The configuration within
layers is violated for both cases. Hence in a way it is easier to
`destroy' an FE phase.

Conversely, starting from an initial random configuration, it is
harder to form the FE phase as particles not just have to couple
together, they have to all reside on one of the layers. This can only
happen at low enough $T$. Thus we may argue that the FE phase is the
dominant phase only at large enough repulsive interlayer couplings
under the drive.

\section{LONG-LIVED TRANSIENTS}

When investigating the transition of the FE to S phase (first order
due to a discontinuity in the structure factor versus $T$ plot),
several {\bf transient} phases are observed. They appear to be the
`local minimum' solutions of an `optimisation problem' in which the S
phase is the best `solution', i.e. configuration of lowest `free
energy' satisfying the parameters of the system.  The new phases
observed are composed of from 2 up to 4 or 5 vertical bands, compared
to the S phase which only has one band. These are dominant at the
comparatively low $T$ for the FE-S transition, whereas we can find the
S phase again at moderate $T$. In fact, these multi-banded structures
had been reported in an Anisotropic Lattice Gas Automata proposed by
Marro et.al., only recently \cite{marro:ALGA}.  In their case, they
have a single lattice gas system evolving not under the Metropolis
rate but automata rules.

\begin{figure}
\resizebox{10cm}{!}{\includegraphics{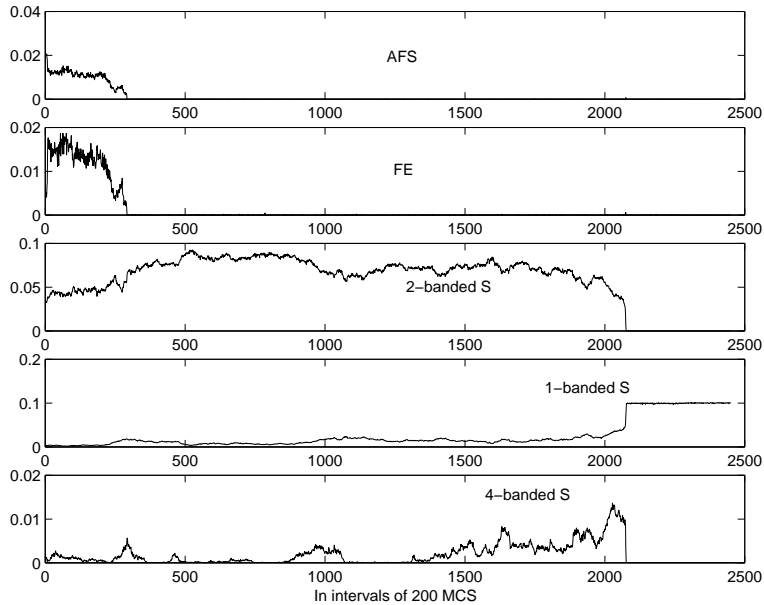}}
\caption{Example illustrating the 2-banded S as a transient to the
final 1-banded S phase.  This is typical of systems with large
$J_2/J_1$ ratios and well inside the intrusion region at negative
$J_3/J_1$, with temperatures well above the S-FE transition line but
less than those values which give very fast convergence to the
single-banded S phase.  The entries in the power spectrum monitored
are: (0,1,1) for AFS, (0,0,1) for FE, (0,2,0) for 2-banded S, (0,1,0)
for 1-banded S and (0,4,0) for the 4-banded S phase.  The run-time
averages taken produce the structure factors for each case.  Note that
the maximum values for $|\tilde{n}(0,2,0)|^2$ and
$|\tilde{n}(0,1,0)|^2$ entries are both 0.1016 for the case of $L$ =
32.}
\label{fig:strc_2-bandedS}
\end{figure}

The $n$-banded S phases are seen to give way to the 2-banded phase as
$T$ increases. For certain runs at moderate $T$, the latter is even
seen to ``evolve'' into the single-banded phase during a long enough
simulation run ( $ > 3 \times 10^5$ MCS). This observation lends
further evidence that the $n$-banded phases are the ``local minima",
from which we could reach the ``global minimum" with an increase in
$T$ or a longer run (implying greater chances given to the
system). See Figure~\ref{fig:strc_2-bandedS}.

% 5 spectrum plots from s_avmf.m

Here, we can also speculate that the cause of the emergence of
$n$-banded S phases is the larger coupling $J_2$. From
\cite{marro:ALGA}, the $n$-banded S phases were obtained with a
setting of 0.9 for a parameter $b$ in their model, with $b \in[0,1]$.
If $b > 1/2$, it implies that there exists a tendency for particles to
approach each other in the transverse direction to the driving
field. For $b < 1/2$, it represents a tendency for particles to
separate from each other. Thus we can see that $b=0.9$ has a similar
effect to a large $J_2$ in our case!  This realization implies that
the $n$- to single-banded S phase transition is a real phenomena in
DDS as it can be produced by different models.

The transients were not reported in Hill's work, probably because the
ratio $\alpha$ is 1. Only when the coupling in the transverse
direction to the drive increases far beyond one can these transients
be observed. These are made more stable by the larger $J_2$. In a way,
the increase of $J_2$ has the effect of ``stretching out" the system
dynamics, making otherwise short or nonexistent transient phenomena
emerge.

\begin{figure}[h]
% structure factor plots.
\resizebox{8cm}{!}{\includegraphics{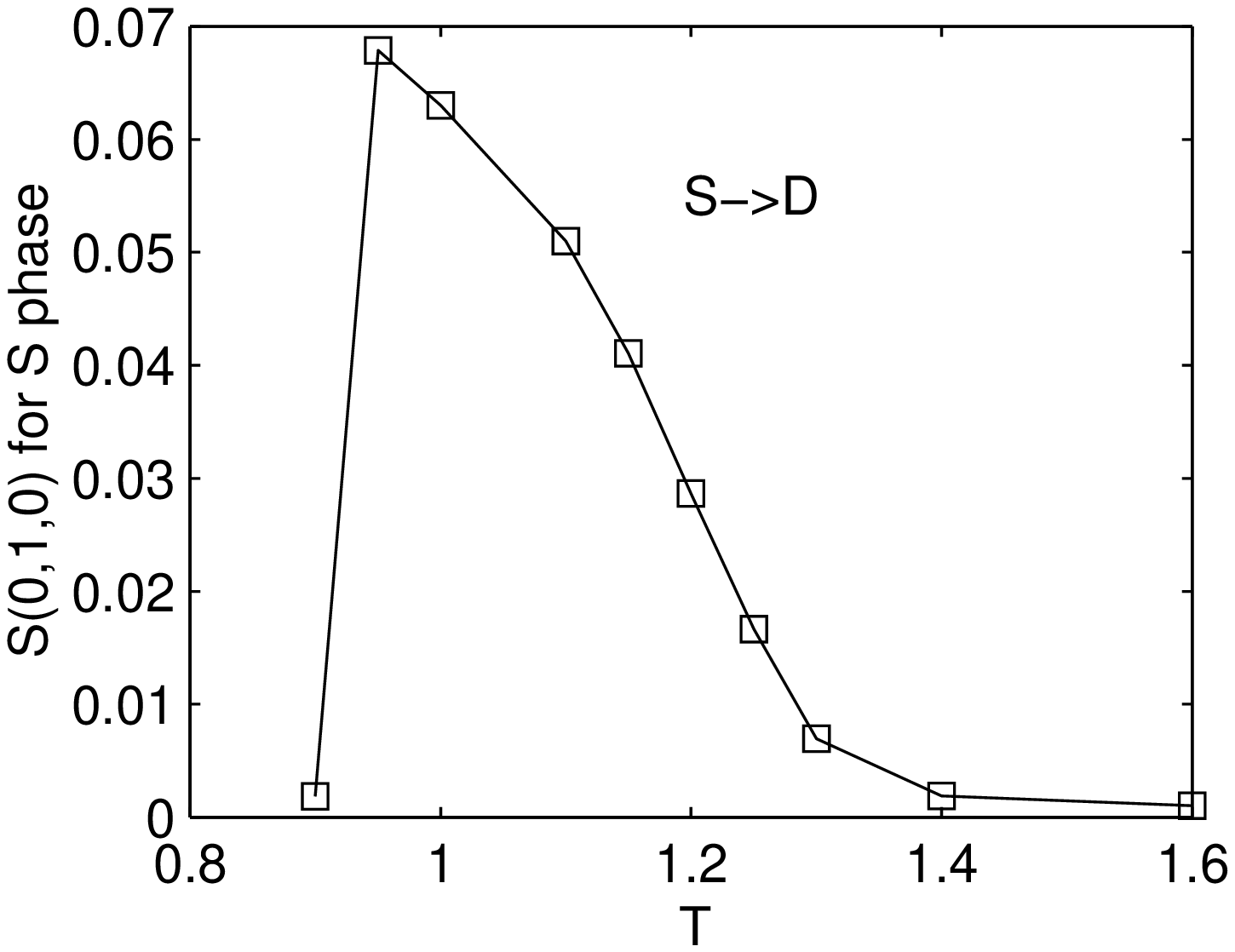}}
\resizebox{8cm}{!}{\includegraphics{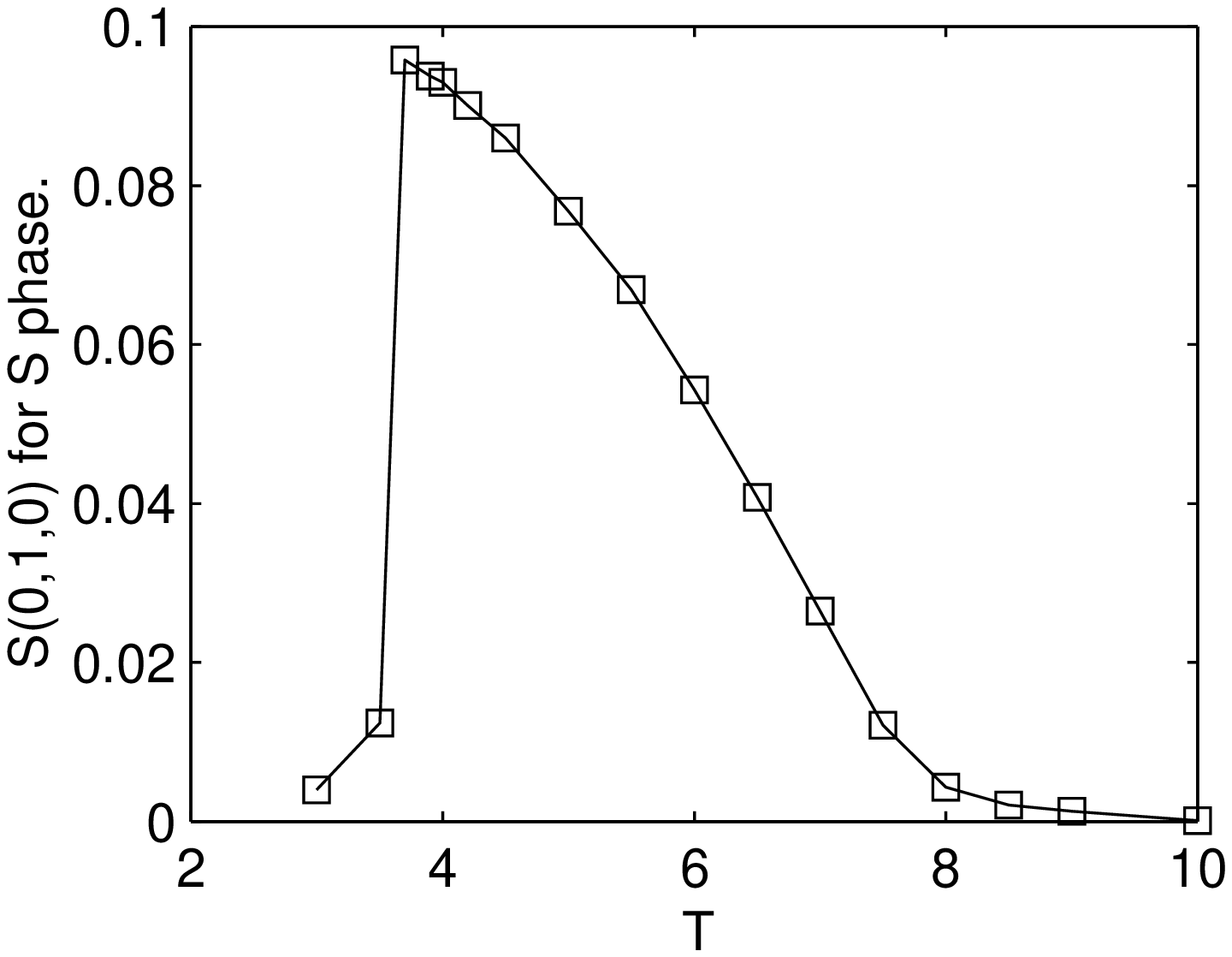}}
\caption{Plots of S(0,1,0) with temperature. The left plot is for $J_{3}= -0.05$ at $J_{2} = 1$ while the right
is at $J_2 = 10$. System size, $L = 32$ and driven at $E_x = 25$.}
\label{fig:strc_fc_-005+1(10)_d}
\end{figure}

Besides making transients longer, the transition to disorder is also
lengthened for systems with larger $J_2$.  This is related to a larger
$T_c$ for D-S transitions. If we plot the structure factors for $J_2$
= 1 and 10, the same shape is observed for both plots but the
temperature range is about 10 times larger for $J_2$ = 10.  Please
refer to Figure \ref{fig:strc_fc_-005+1(10)_d} for the `shark's fin'
plots.  This `glassy' behaviour is speculated to be also due to the
larger $J_2/J_1$ ratio.  Observe the first order transition at the low
temperature end and the second order transition at higher
temperatures. The first order transition is due to a FE-S transition
for $J_2$ = 1 whereas it is for a $n$-banded S to 1-banded S for the
$J_2$ = 10 case.

Finally, some words about obtaining the FE-S first order transition
line. For $J_2 > J_1$, the FE phase is seldom observed inside the
`triangular' region. Instead, either the AFS phase or a sort of {\bf
mixed phase} having both AFS and $n$-banded S characteristics are
observed. This led to the $n$-banded S phase at higher $T$. Thus we
are seeing another transient configuration. Their appearance
effectively fuzzed out the first order transition line and so a
heuristic approach has to be taken. We simply take the smallest $T$
which gives a n-banded S phase as an estimate of $T_c$.

\section{CRITICAL EXPONENT DETERMINATION}

Critical exponents, unlike the critical temperatures which depend very
much on the details of the model system, only depend on a few
specifications of the system. For models with short-range
interactions, like in our case, these are simply the dimensionality of
space and the symmetry of the order parameter.  All models with the
same exponents belong to the same class, of which the Ising
universality class is the most common, labelled by the simplest
member.

In the paper by Hill, of which the present work is based, it was
mentioned that work was in progress to identify the universal
properties of the D-FE transition in our model. Though no concrete
results were published, we worked under the hypothesis that it is
Ising due to the wide applicability of the class, unless proven
otherwise.  The strategy we adopted was to either prove or disprove
the Ising class hypothesis.

The current status of knowledge in the field was that for a KLS model,
it belongs to the DLG class. If we remove the drive, it is reduced to
an Ising model due to the equivalence between spin and lattice gas
systems.  For a bilayered structure with two KLS systems stacked on
top of each other but uncoupled, the model exhibits two phase
transitions of which D-S is DLG and D-FE is Ising. Finally, removing
the drive for the above system and we should expect two Ising
systems. The effect of adding coupling to a driven system is currently
being studied.

We tried to determine the universality class for the D-FE transition
under a finite but large drive. Working under the hypothesis that the
system is still Ising, we computed the quantity $\gamma/\nu$ to see if
the Ising value of $7 \over 4$ can be obtained. This is done by
assuming the finite-size scaling relation,
\begin{equation}
\chi_{max}(L) \propto L^{\gamma/\nu}.
\end{equation}
Hence, by getting good estimates of the susceptibility peak values for
various system sizes, we can obtain an estimate for the ratio
$\gamma/\nu$.

Before we proceed, we would like to say something on the critical
exponents.  The exponent $\gamma$ controls the divergence of the
susceptibility function near the critical point, as in the power law,
\begin{equation}
\chi \propto \left| T - T_c \right|^{-\gamma}.
\end{equation}

The value for the 2-D Ising model is 7/4. As for the exponent $\nu$,
it is called the correlation length exponent and takes on the value 1
for the Ising model. It controls how the correlation length diverges
near criticality.

Let us outline the tactic we used. For a given $J_2$ setting, we
attempt to obtain estimates of $\gamma/\nu$ as well as the individual
exponents $\gamma$ and $\nu$ for representative $J_3$ values, namely
$-1$, $-5$ and $-10$. To do this, we require more detailed
susceptibility plots especially for the region near the peak, where
systems with $T$ values differing only in the 3rd decimal place are
investigated. Data points close to the peak are fitted with a
least-squares quadratic polynomial and the maximum value as well as
its location determined. These are the $\chi_{max}(L)$ and
$T_{peak}(L)$ we desire. By repeating the procedure for system sizes
$L$ = 32, 64 and 128, we could plot $T_{peak}$ vs $L$ with a guess for
$\nu$ to obtain $T_c$.  Naturally, $\nu$ = 1.0 is chosen to test our
hypothesis.

By plotting $\log \chi_{max}$ vs $\log L$, the gradient of the
least-squares fit straight line gives the ratio $\gamma/\nu$.  This
value is then used in the $\log(\chi L^{-\gamma/\nu})$ versus $\log(
|T-T_c|L^{-1/\nu})$ plot with $\nu$ set to 1.0. (This plot shall also
be called ``scaling plot" for short.)  With this we can check to see
if the derived quantities gives good ``data collapse'', which is
expected if the scaling relations are satisfied. From the plot, the
slopes of the two best-fit straight lines is expected to give us the
exponent $\gamma$. In other words, if the simulation data fits the
finite-size scaling theory well, we should obtain two branches which
are well-fitted by straight lines with the same slope, characterising
the same power law behaviour of the $\chi$ values as the critical
point is approached.

Before we present our data and make any conclusions with regards to
the universality class of our DDS, we would like to present the
computed heat capacity values from our model and compare with the
exact Ising results. It is clear that under no drive and without any
interlayer interactions, we have essentially two separate 2-D Ising
systems.  Hence, by setting $J_3=0$ and $J_2=J_1$, simulation runs are
performed for system sizes $L$ = 4, 8 and 16. Starting with the
definition of the heat capacity for the 2-D Ising model, we derived
the equation that relates the particle Hamiltonian for our model to
that of the 2-D Ising spin system given below,
\begin{equation}
c_{\sigma} = {L^2 \over {k_B T^2}} [ \langle e_{\sigma}^2 \rangle   -  {\langle e_{\sigma} \rangle}^2],
\end{equation}
with $e_{\sigma} = 4( H/{2 L^2} ) + 2J_1$ and $T$ is given
in the spin language, i.e. $T_c = 2.26$.  As a reminder, $H$ is the
total energy of the lattice gas system.

Table \ref{table:heat_cap} gives the numerical results as compared to
Ising.  The length of runs taken is $10^7$ MCS, which is
achievable for such small systems. The first $5 \times 10^4$ MCS are
skipped to avoid transient phenomena.  Note that the results are for
the equivalent spin system in order to compare with the Ising model
and that double precision floating point arithmetic is used, with 5
separate runtime averages from random initial conditions used for
computing the average heat capacity.

% table of heat capacity values.
\begin{table}
\begin{tabular}{|l|r|r|r|}   \hline
$L$ &  Average $c_\sigma$  &  Standard deviation & Exact (Ising) \\ \hline\hline
4     &  0.78328  & 0.0038 & 0.78327 \\ \hline
8     &  1.1468  & 0.0078 & 1.1456\\ \hline
16     &  1.5050  & 0.0067 & 1.4987 \\ \hline
\end{tabular}
\caption{Comparison of the heat capacities computed from the model with the exact values from Ising.}
\label{table:heat_cap}
\end{table}

As the Table shows, we have good agreement with the exact Ising values. This provides evidence that our model is
implemented correctly.

As a first application of the method outlined above to estimate the
ratio $\gamma/\nu$, we investigated the universality properties for a
decoupled, undriven and isotropic lattice gas, essentially expecting
to see Ising behaviour.  With simulation runs of $5\times 10^5$ MCS,
the $T_{peak}(L)$'s for $L$ = 4, 8, 16 and 32 are estimated to be
1.389, 1.165, 1.070 and 1.036 respectively. Theoretical arguments give
$T_c$ as 1.0. Hence we see that finite-size effects are indeed at work
to shift $T_{peak}$ values further from the true $T_c$ as $L$
decreases.

With these, plots of $T_{peak}$ vs $L^{-1/\nu}$ as well as that of
$\log{\chi_{max}}$ vs $\log{L}$ are made.  It was found that if we do
not include the $L = 4$ data, the value of $T_c$ obtained assuming
$\nu$ = 1.0 is 0.9885 compared to 0.9729 if we do. This is evident
that data from the $L$ = 4 system is too small.

The estimates of the peak heights are 0.2347, 0.8213, 2.9207 and
9.3172 respectively. Only the last three values are used in the latter
plot, from which the gradient gives an estimate of 1.7520 for the
ratio $\gamma/\nu$.  This values has a relative error of only 0.11$\%$
as compared to the Ising value of 1.75!  Assuming exponent $\gamma$ to
be 1.75, we obtained the experimentally obtained $\nu$ value of 0.9989
(very close to 1.0) with $T_c$ then obtained as 0.9886.  Hence, both
cases give $T_c$ very close to the expected value of 1.0.

\begin{figure}
\resizebox{10cm}{!}{\includegraphics{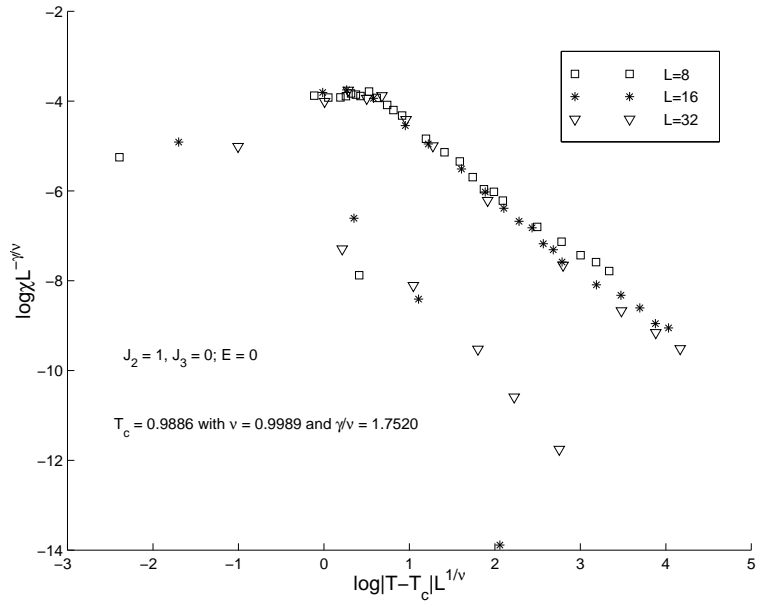}}
\caption{``Data collapse" plot for the case of $J_2 = 1$, $J_3 = 0$ and $E = 0$ using all experimental values.}
\label{fig:dsp_j30j21expt}
\end{figure}

From the scaling plot, Figure \ref{fig:dsp_j30j21expt}, the value of
$\gamma$ is estimated from the slope of the least-squares line fitting
the linear portion of the upper data points. This turns out to be
1.7273 for the cases of both $[\nu=1.0, T_c = 0.9885]$ and $[0.9989,
0.9886]$. As the percentage error of this value from the assumed value
of 1.75 is only $1.30\%$, we conclude that the undriven, decoupled
bilayer system is indeed Ising in nature.  This is the expected result
as we have, in fact, two independent Ising systems.

\begin{figure}
\resizebox{10cm}{!}{\includegraphics{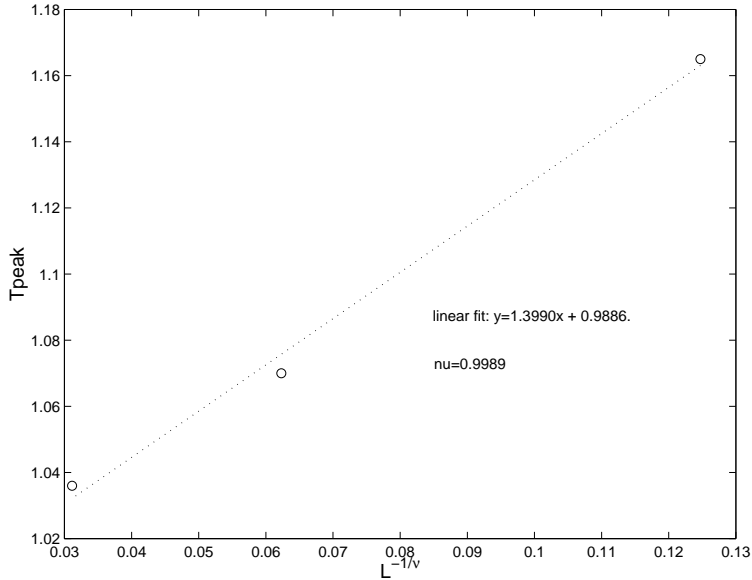}}
\caption{Plot of $T_{peak}$ vs $L^{-1/\nu}$ for the undriven,
decoupled and isotropic case. The critical temperature obtained is
0.9886, very close to the expected value of 1.0.}
\label{fig:tpeakl_dc}
\end{figure} 

With this much groundwork done, we can proceed to the new
findings. Due to time and resource constraints, only $J_2=1$ and a
portion of the $J_2=2$ FE-D phase space is explored to determine the
universality class.  As a rough guide, the CPU time spent on this
portion of the paper was about 1800 hours (an underestimate) for a
Digital Alpha processor running at 600MHz. Typical running times: 1
hour for $L=32$, 5 hours for $L=64$ and 24 hours for $L=128$, all with a run
length of $1 \times 10^6$ MCS. These resource hungry tasks are
completed thanks to a cluster of 30 Compaq Personal Workstations at
the Department of Computational Science, NUS. Running under the Condor
batch submission system developed at the University of Wiscosin, USA,
which enables the simultaneous running of up to 10 jobs, all runs were
started from the same initial (randomly) half-filled configuration but
at different temperatures.  Our results seem to indicate a deviation
from Ising when the system is placed under the large but finite drive.

\begin{figure}
\resizebox{8cm}{!}{\includegraphics{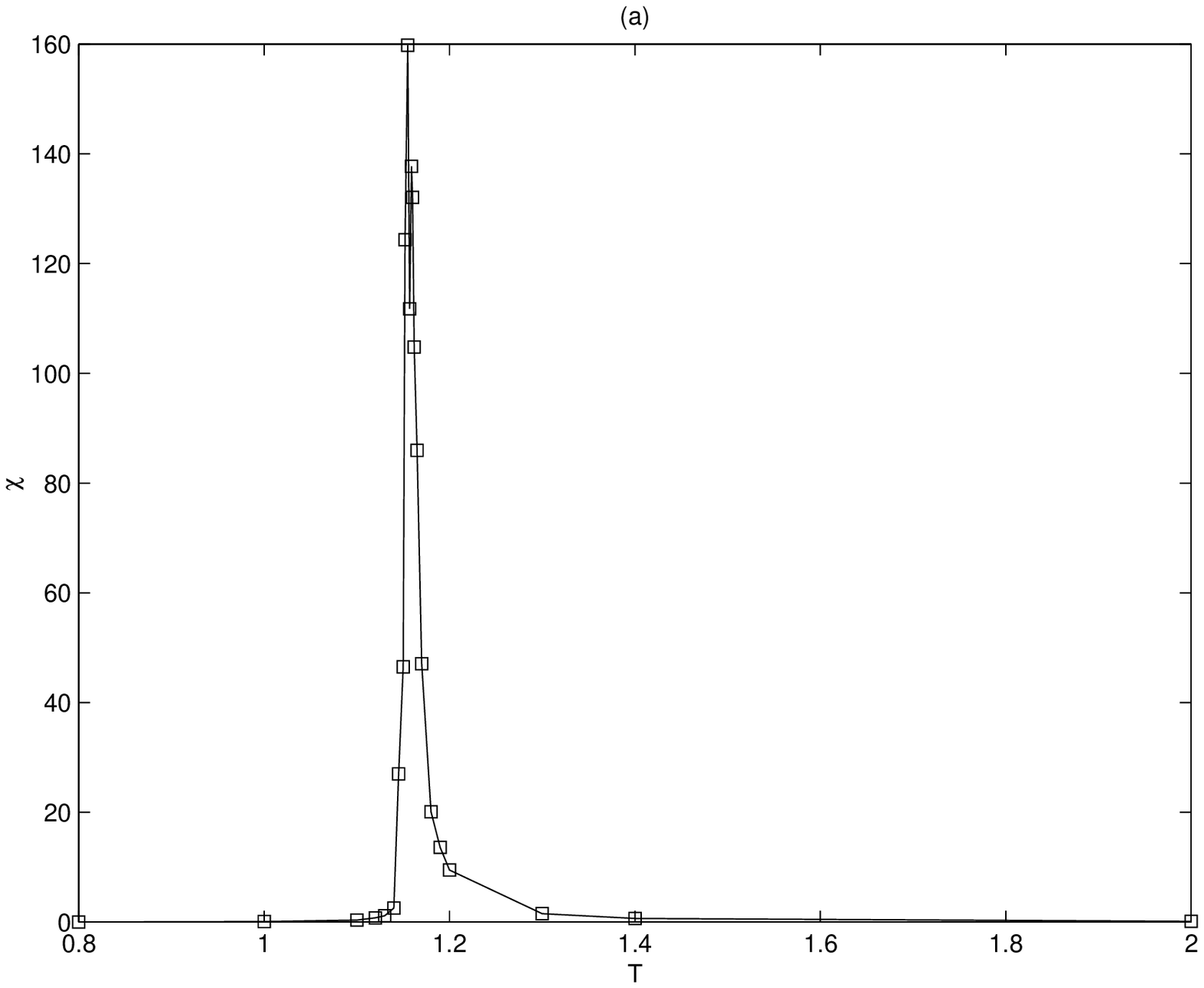}}
\resizebox{8cm}{!}{\includegraphics{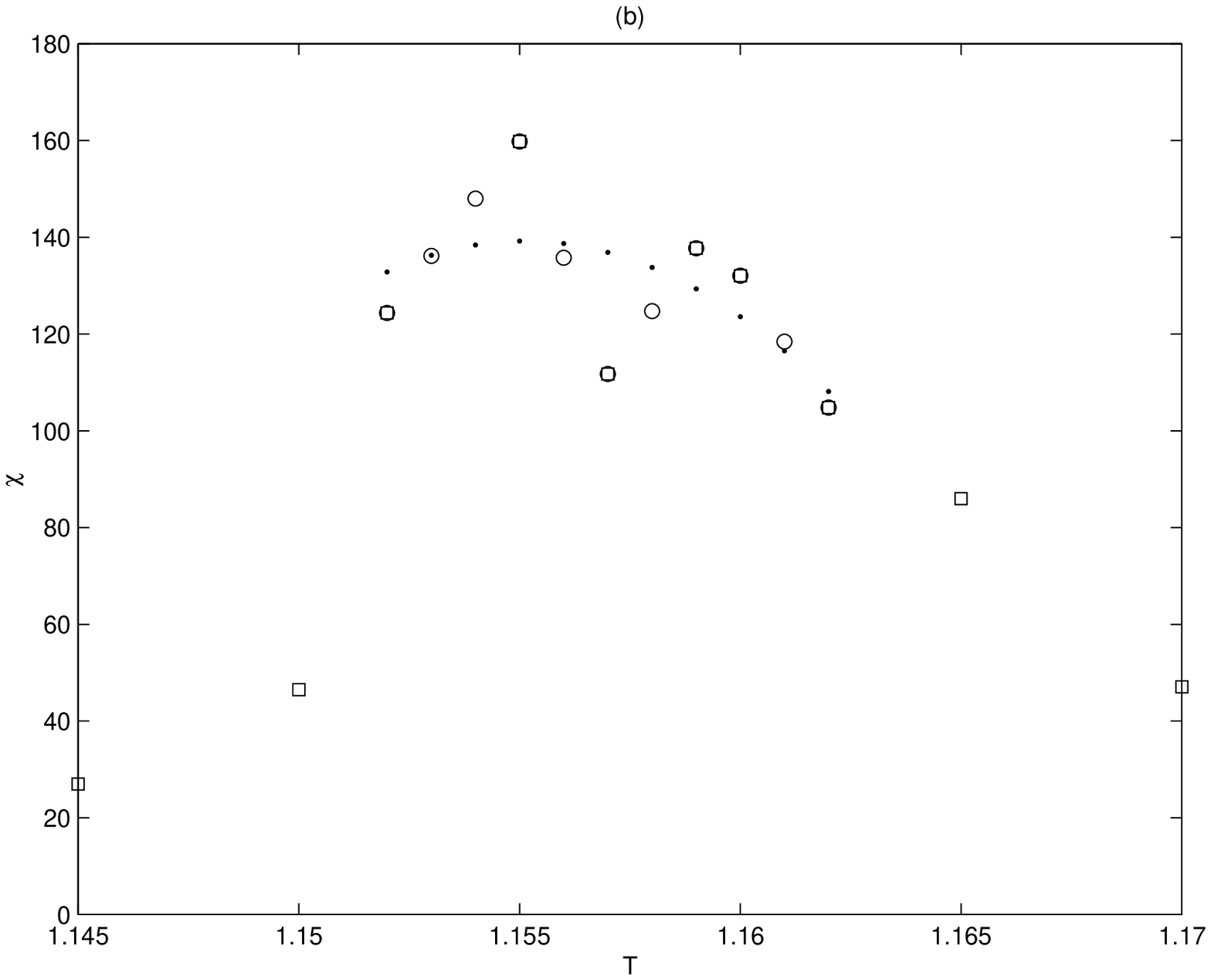}}
\caption{Plots of $\chi$ against $T$ for $J_2 = 1$ and $J_3 = -1$, with (a) being the full plot and
(b) giving a zoomed-in picture. The simulation data are shown as squares while the dots in the second figure 
represent the attempt to fit a quadratic curve through the interpolated values. The circles are artificial data points
generated by linear interpolation between the experimental data in order to improve the eventual quadratic fit.}
\label{fig:chi_plot_sample}
\end{figure}

First of all, we would like to give a figure depicting the problems we
faced in the determination of the peaks for the susceptibility
plots. See Figure \ref{fig:chi_plot_sample} for the plots of the peak
as well as a zoomed-in portion where the $\chi_{max}$ and $T_{peak}$
values are estimated through a quadratic fit.

As shown in Figure \ref{fig:chi_plot_sample}(b), data points about the
peak are sort of jagged. In theory, the susceptibility values do not
grow infinitely large due to the finite size of the model system. They
should be ``rounded" at the top due to the finite system size, over
the range of temperatures for which the correlation length $\xi$ is
close to $L$. In practice, data points are scattered about some
fitting quadratic polynomial. This observation could be due to
critical slowing down of the dynamics near criticality due to
divergence of $\xi$. Hence, we need an estimate of how well the
polynomial fits the data values, thus giving us an estimate of the
error associated with the maximum $\chi$ value obtained via the fit.

We attempt to associate an error with the estimate of $\chi_{max}$
through the following heuristic approach.  From the set of data points
about the observed peak of the function, a linear interpolation is
made to obtain more points. The difference between these pseudo data
points and those from the parabolic fit to the chosen interval is
denoted by $\epsilon$ ($= y_{data} - y_{fit}$). Due to plotting
limitations, artificial data points are introduced through a linear
interpolation which should preserve the original nature of the data
and thus $\epsilon$ can only be close to zero in the best cases.  We
next compute the variance of the set of $\epsilon$ values as
$var(\epsilon) = \langle \epsilon^2 \rangle$ and take the standard
deviation, $\sigma(\epsilon) = \sqrt{var(\epsilon) \over (n-1)}$ as an
estimate of the error in $\chi_{max}$. This gives us a gauge as to the
spread of the ``errors" when the data points are fitted by a
least-squares degree 2 polynomial.  However, this estimate does not
tell us how far our estimate is from the true $\chi_{max}$ for the set
of parameters, as effects like critical slowing down may be present to
alter the observed peak height.

\begin{figure}
\resizebox{8cm}{!}{\includegraphics{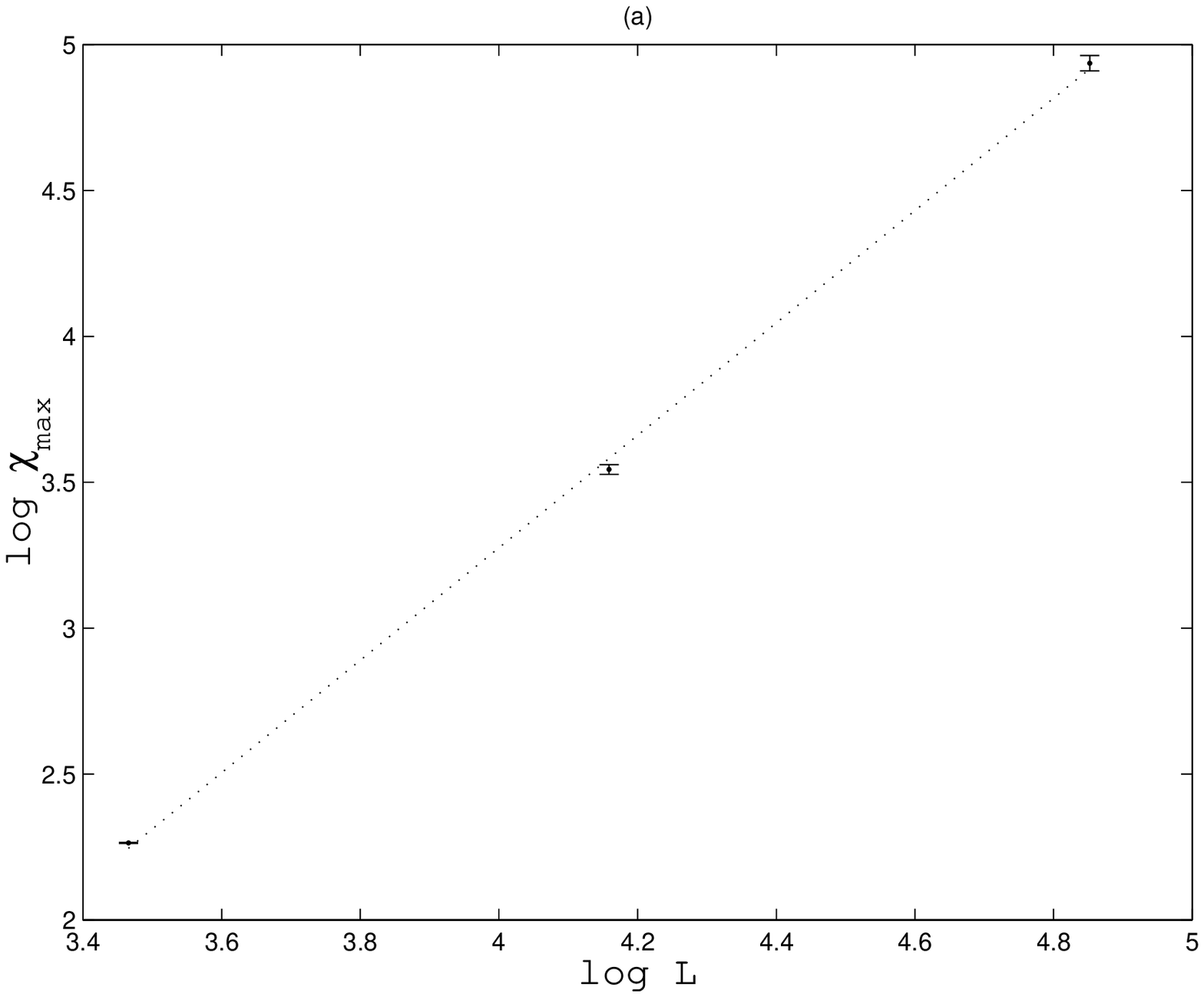}}%
\resizebox{8cm}{!}{\includegraphics{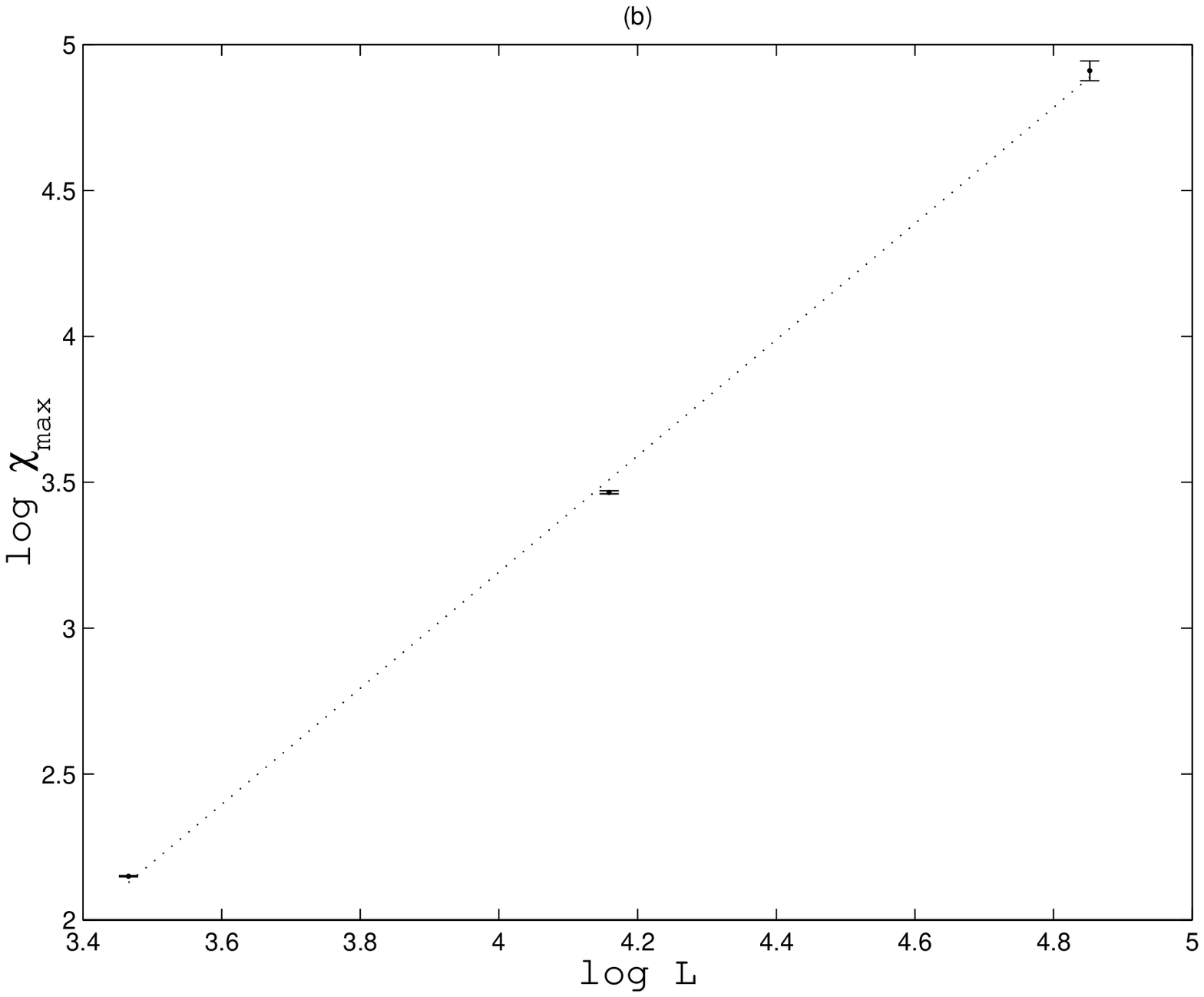}}
\resizebox{8cm}{!}{\includegraphics{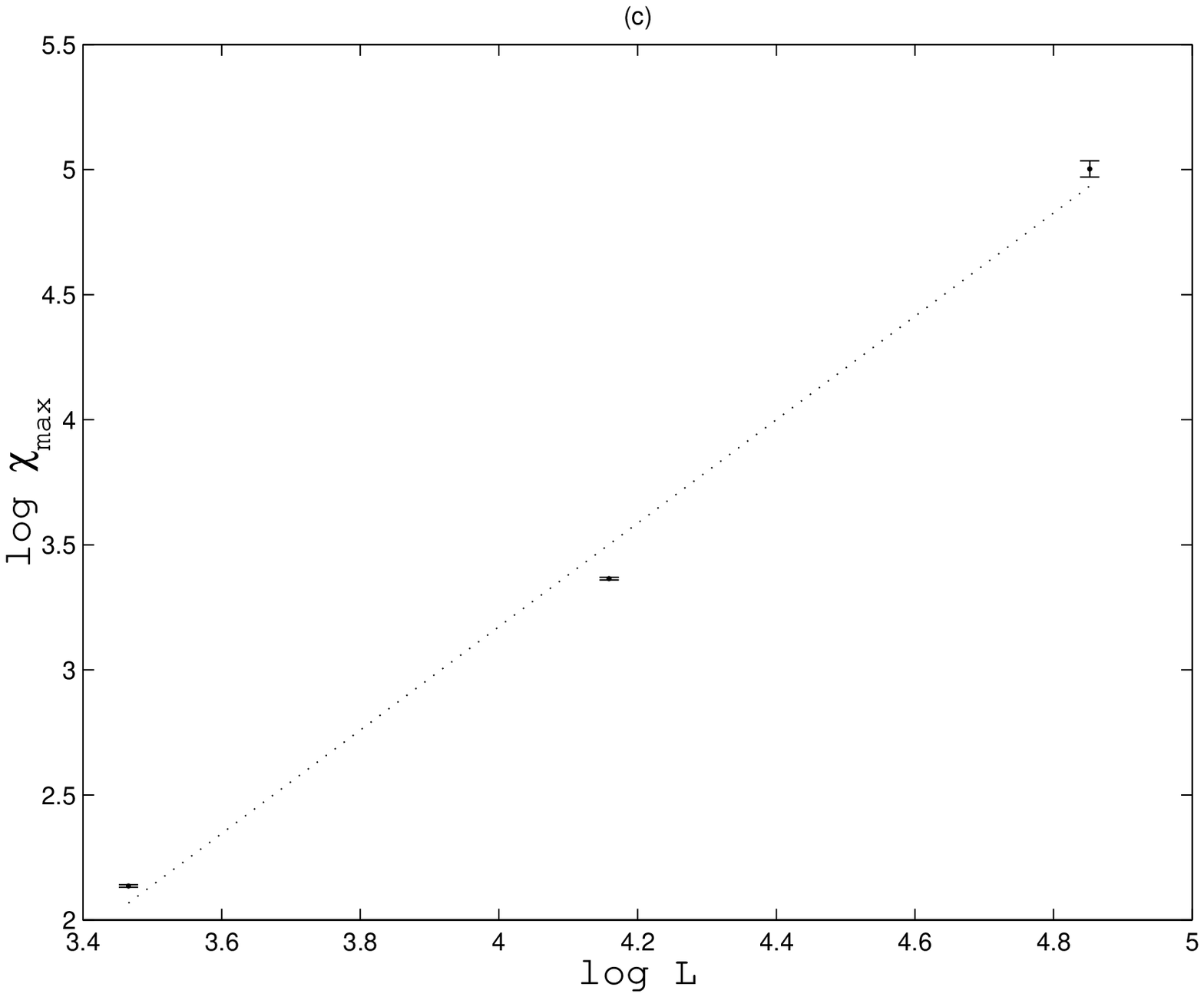}}%
\resizebox{8cm}{!}{\includegraphics{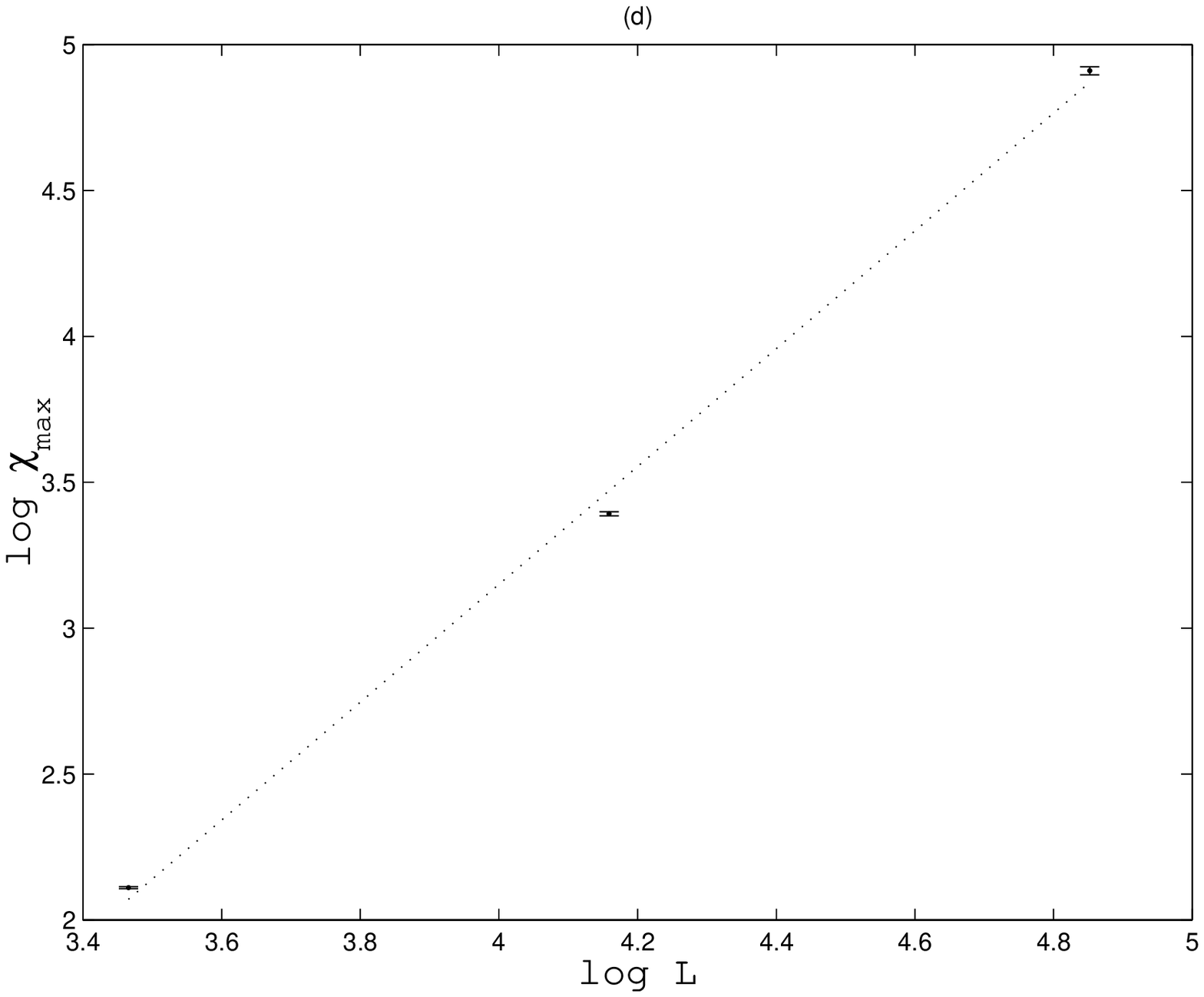}}
\caption{Plots giving a feel of the errors associated with each
$\chi_{max}$ value obtained.  Each plot is for a different set of
coupling, with $J_2 = 1$ and $J_3$ being $-1$ for plot (a), $-2$ for
(b), $-5$ for (c) and $-10$ for (d). Error bars are computed as
explained in the text.}
\label{fig:gammaonu}
\end{figure}

In Figure \ref{fig:gammaonu}, we plotted the log-log plots of
$\chi_{max}$ data versus the system sizes $L$ investigated. The error
bars plotted represent twice the propagated errors in
$log(\chi_{max})$ which is the error in $\chi_{max}$ divided by
$\chi_{max}$. It is observed that in general the errors associated
with the largest system size of $L=128$ is larger, but not large
enough to cause a significant variation in the slopes.

% gammaonu plots giving estimates of gamma/nu.

Table \ref{table:gamma_o_nu's} lists the estimates for the ratio
$\gamma / \nu$ based on taking the ratio of $\log(\chi_2/\chi_1)$ over
$ \log(L_2/L_1)$. Here $\chi_1$ is the short form of $\chi_{max 1}$
for system size $L_1$.  Listed are the values for different ratios
$L_2/L_1$ as well as the propagated errors in $\gamma / \nu$, which is
$\delta(\gamma/\nu) = {1 \over {\log(L_2/L_1)}} [ \sigma_2
/\chi_2 + \sigma_1/\chi_1 ]$, where the natural logarithm is
taken.

% table of gamma over nu values.
\begin{table}
\begin{tabular}{|l|r|r|r|r|r|}   \hline
$J_3/J_1$ & $L_2/L_1$ & $\gamma/\nu$ & $\delta(\gamma/\nu)$ & ${\gamma \over \nu} - \delta({\gamma \over \nu})$ &
${\gamma \over \nu} + \delta({\gamma \over \nu})$ \\ \hline
-1    & 64/32     & 1.846       & 0.026    &  1.819  &  1.872 \\ \cline{2-6}
      & 128/64    & 2.009       & 0.062    &  1.946  &  2.071 \\ \cline{2-6}
      & 128/32    & 1.927       & 0.020    &  1.907  &  1.947 \\ \hline \hline

-2    & 64/32     & 1.898       & 0.011    &  1.887  &  1.909 \\ \cline{2-6}
      & 128/64    & 2.084       & 0.056    &  2.029  &  2.140 \\ \cline{2-6}
      & 128/32    & 1.991       & 0.004    &  1.987  &  1.995  \\ \hline \hline

-5    & 64/32     & 1.774       & 0.015    &  1.759  &  1.789 \\ \cline{2-6}
      & 128/64    & 2.363       & 0.053    &  2.309  &  2.416 \\ \cline{2-6}
      & 128/32    & 2.068       & 0.027    &  2.042  &  2.095 \\ \hline \hline

-10   & 64/32     & 1.848       & 0.016    &  1.832  &  1.863 \\ \cline{2-6}
      & 128/64    & 2.191       & 0.030    &  2.161  &  2.221 \\ \cline{2-6}
      & 128/32    & 2.019       & 0.013    &  2.007  &  2.032 \\ \hline
\end{tabular}
\caption{Ratios of $\gamma/\nu$ computed from various scenarios, with
the associated errors. Also included are the intervals for various
estimates of the ratios.}
\label{table:gamma_o_nu's}
\end{table}

From the Table, it is clear that all intervals for $\gamma/\nu$
computed do not include the value 1.75. An important observation is
that for ratios computed using the $L=128$ data, a value greater than
2.0 can be obtained!  This data does not fit into our scheme of things
so far which places a limit that $\gamma/\nu$ is less than 2.

If we take the upper bound of the ratio $\gamma/\nu$ to be 2.0, it
would mean that the data points for $L=128$ may be inaccurate. As the
errors computed could not explain the discrepancy, it was suspected
that critical slowing down is quite severe in such a large system size
and that 1 million MCS taken was not sufficient for the system to
reach the true steady state. If this is indeed the case, then the data
for $L$=32 and 64 should be more trustworthy.  But their intervals
also do not include 1.75. Thus it is concluded that we observe here a
significant deviation from the Ising value of 1.75 for the ratio of
the exponents.

\begin{figure}
\resizebox{10cm}{!}{\includegraphics{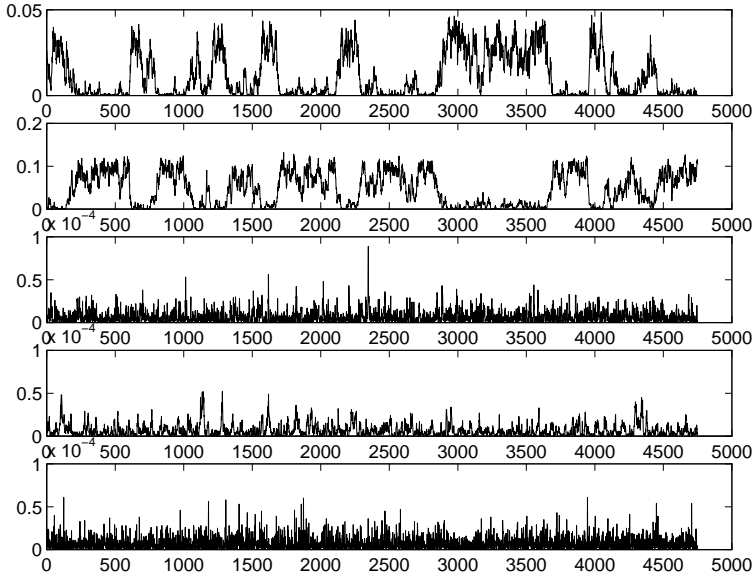}}
\caption{Fluctuations of $|\tilde{n}|^2$ with time near $T_c$ for the
driven system. Parameters: $J_3 = -10$, $J_2=1$, $L=128$ and
$T=1.2405$.  The quantities plotted on the y-axes are (from top to
bottom): $|\tilde{n}(0,1,1)|^2$, $|\tilde{n}(0,0,1)|^2$,
$|\tilde{n}(0,2,0)|^2$ , $|\tilde{n}(0,1,0)|^2$ and
$|\tilde{n}(0,4,0)|^2$, as in Fig.\ref{fig:strc_2-bandedS}.  The
horizontal time axis is in units of 200 MCS.  Note that as the value
in the second row (FE) increases, that of the first row (AFS) decreases and
vice-versa.}
\label{fig:fluctuate}
\end{figure}

\begin{figure}
\resizebox{10cm}{!}{\includegraphics{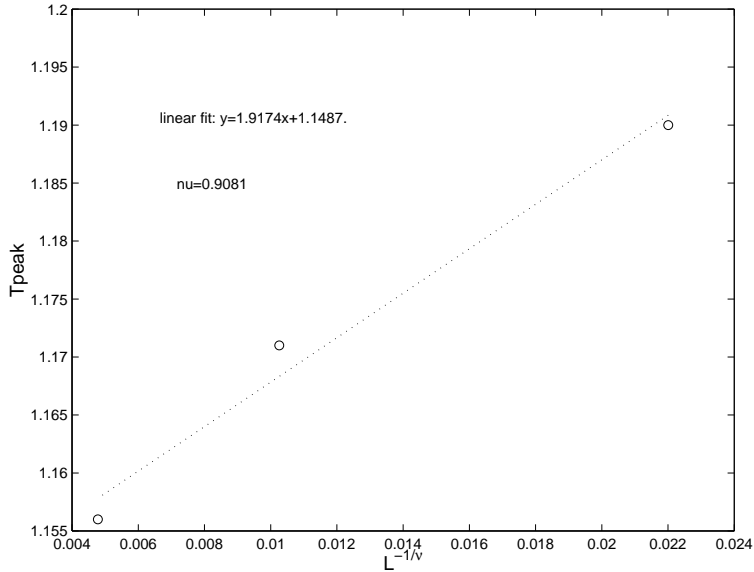}}
\caption{Plot of $T_{peak}$ vs $L^{-1/\nu}$ for the driven system with $J_3 = -1$ and $J_2 = 1$.}
\label{fig:tpeakL}
\end{figure}

With the experimental ratios of $\gamma/\nu$, we assumed $\gamma$ to
remain at the Ising 1.75 value and plotted $T_{peak}$ against
$L^{-1/\nu}$ for each setting of coupling strengths investigated.
With $\nu < 1$, or for that matter with $\nu = 1.0$ for Ising systems,
the plots obtained could not be reasonably fitted with least-squares
straight lines.  In fact, all plots seem logarithmic-like. Is this
another signature of a non-Ising system or the existence of two
correlation lengths? We could not provide an answer at this current
stage of research.  In order to proceed, we used a linear fit to
obtain a $T_c$ via extrapolation using the ``experimentally" obtained
value of $\nu$. See Fig. \ref{fig:tpeakL} for a representative plot.

We made ``scaling plots" for the different system sizes for each value
of the parameter $J_3$ investigated. As our susceptibility $\chi$
plots show much similarity with Ising plots, we assumed that the
exponent $\gamma$ which determines the power-law scaling of the $\chi$
plots on either side of the peak to remain Ising, i.e. it has a value
of 1.75. However, this would imply that the exponent $\nu$ is less
than 1.0! Hence we plotted the curves with exponent $\nu$ set to 1.0
as well as the computed value and compared between the plots, besides
observing whether the slopes of the upper and lower best-fit straight
lines give the $\gamma$ value assumed. It was found that the ``Ising"
plots were not consistent in that we do not recover the assumed
$\gamma$ value of 1.75 from the slopes.  There are altogether eight plots
for the four $J_3$ settings we looked at (with $J_2/J_1 = 1$).  We
realised that for consistency, we cannot use the $L$ = 32 and 64 data
to estimate the $\gamma/\nu$ yet deal with all three sets in the
determination of $T_c$ and in the ``scaling plots". For that, we
boldly assume that the ratio of $\gamma/\nu$ in our model is indeed
close to 2.0! This would imply a non-Ising character, where
justification will be presented later.

\begin{figure}
\resizebox{10cm}{!}{\includegraphics{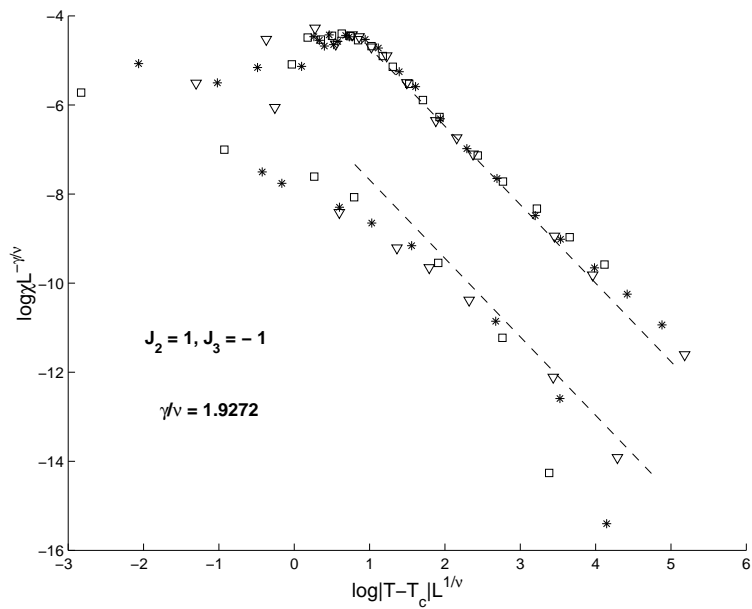}}
\caption{``Data collapse" plot for the case of $J_2 = 1$, $J_3 = -1$
using all experimental values. System sizes used were 32 ($\Box$), 64
($\ast$) and 128 ($\bigtriangledown$). The dotted lines represent
slopes of 1.75.}
\label{fig:dsp_j3-1j21_expt}
\end{figure}

\begin{figure}
\resizebox{10cm}{!}{\includegraphics{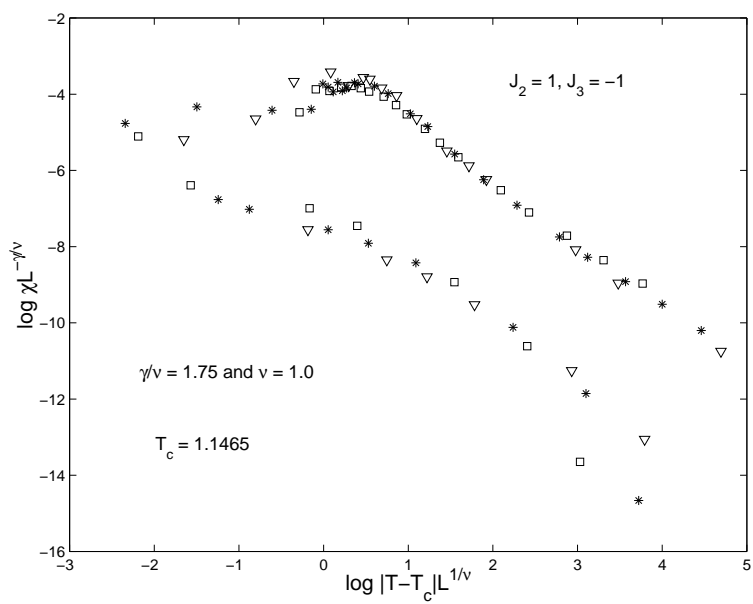}}
\caption{``Data collapse" plot for the case of $J_2 = 1$, $J_3 = -1$
assuming Ising values.}
\label{fig:dsp_j3-1j21nu1}
\end{figure}

From the scaling plots with the experimental values, we observed that
the straight line of slope 1.75 can be fitted through the data points
in the linear regions. Thus, the assumption of $\gamma$ being 1.75 is
consistent with the plots. Further, we observed that the data points
for different system sizes shows signs of scaling behaviour, in that
data points from smaller systems deviate from the perceived linear
region faster. This applies for both the top and bottom branches and
is much due to finite-size effects. Another point to note is the very
short linear regions obtained from the model.  Finally, compare Figure
\ref{fig:dsp_j3-1j21_expt} with Figure \ref{fig:dsp_j3-1j21nu1} where
the exponents assume Ising values. The data collapse near the ``bend"
is not as good as in the former plot.

Similar situations occurred for the other settings of $J_3$, where the
values we sampled ranged from close to the bi-critical point to well in
the region of large repulsive inter-layer potentials. All the slopes
measured are close to the value of 1.75 assumed. Again, collapse is
visually better with the ``all-experimental" cases.

\begin{figure}
\resizebox{10cm}{!}{\includegraphics{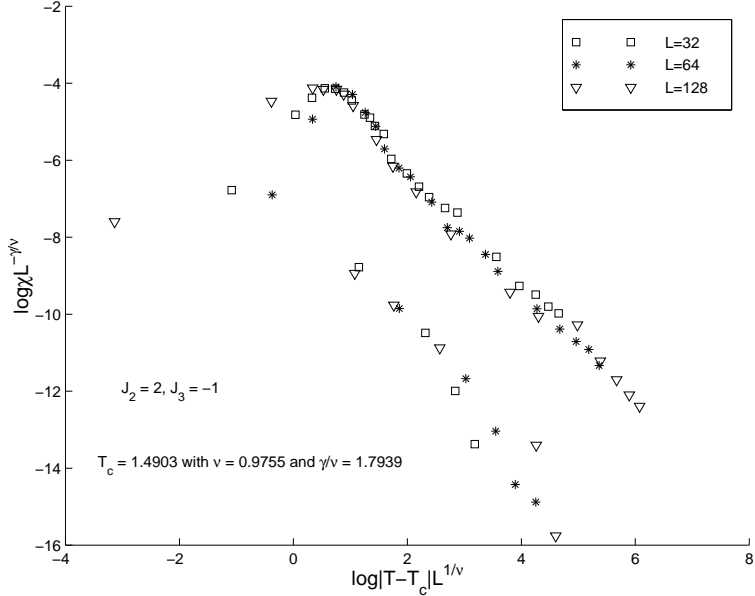}}
\caption{``Data collapse" plot for the case of $J_2 = 2$, $J_3 = -1$
using experimental values.}
\label{fig:dc_j3-1j22_expt}
\end{figure}

\begin{figure}
\resizebox{10cm}{!}{\includegraphics{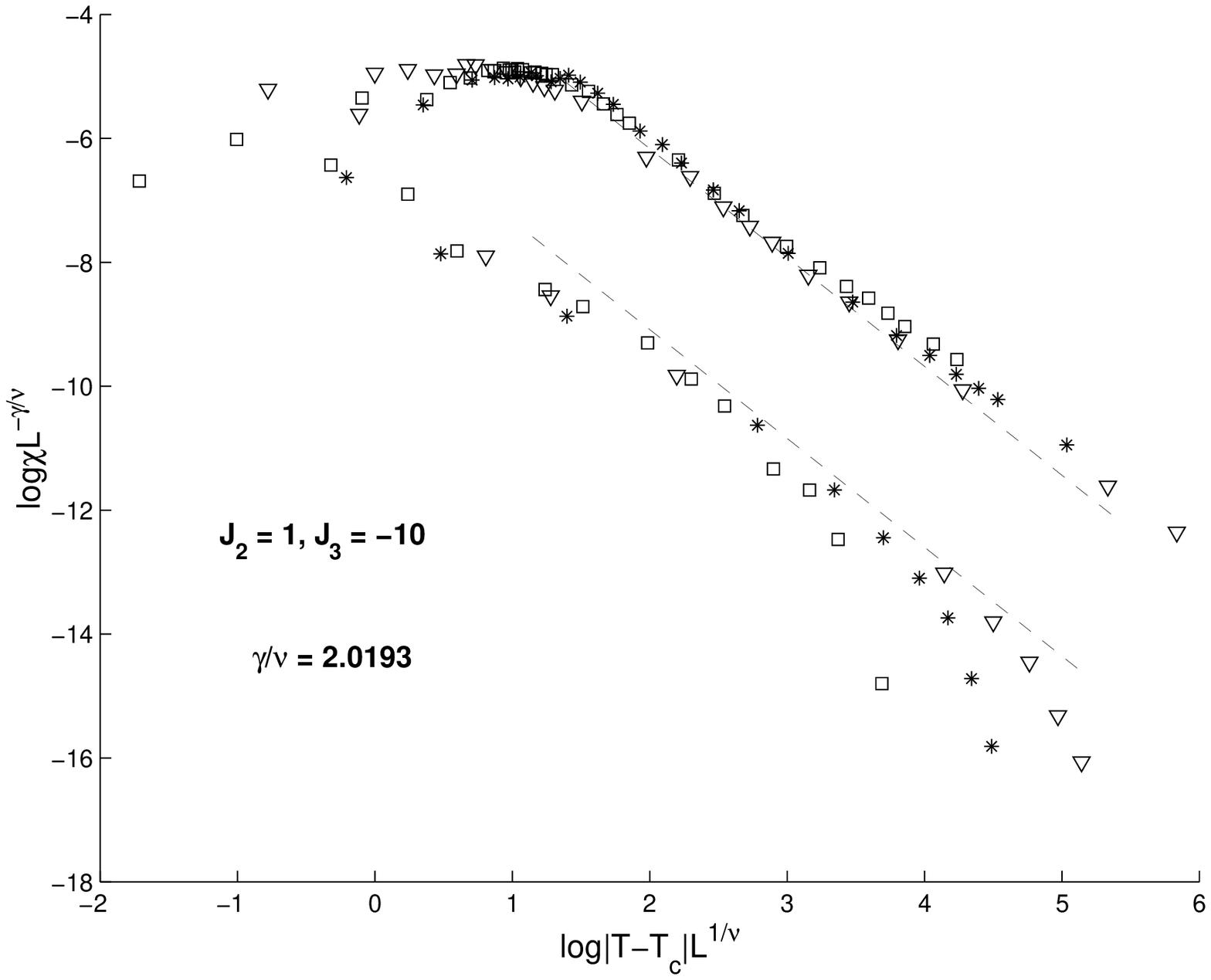}}
\caption{``Data collapse" plot for the case of $J_2 = 1$, $J_3 = -10$
using experimental values.}
\label{fig:dsp_j3-10j21_expt}
\end{figure}

\begin{figure}
\resizebox{10cm}{!}{\includegraphics{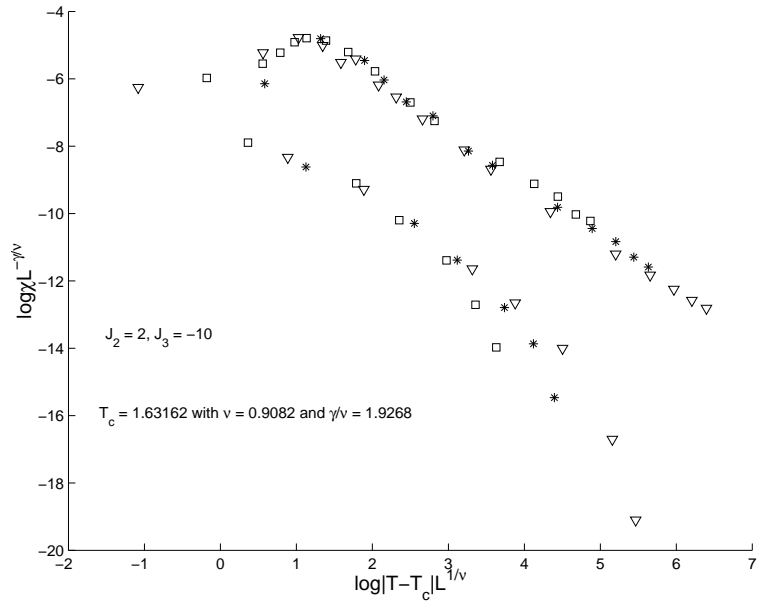}}
\caption{``Data collapse" plot for the case of $J_2 = 2$, $J_3 = -10$
using experimental values.}
\label{fig:dc_j3-10j22_expt}
\end{figure}

We also moved on to look into the case where $J_2 / J_1$ is larger
than 1. Compare Fig. \ref{fig:dsp_j3-1j21_expt} with
Fig.\ref{fig:dc_j3-1j22_expt} and Fig. \ref{fig:dsp_j3-10j21_expt}
with Fig. \ref{fig:dc_j3-10j22_expt}.  It is not difficult to observe
that the data collapse is not as good in the case of $J_2 = 2$.  Does
this imply that the deviation from Ising is more severe for this case?
It is hard to make any statements as current knowledge indicates that
intra-layer couplings are not expected to affect the universality
class of the model system. However, although $J_3 = -10$ gave us a
$\gamma/\nu$ ratio of 1.9268, that of $J_3 = -1$ is only 1.7939, which
is still a puzzle. As the susceptibility plots from $J_2=2$ is similar
in nature to those from $J_2=1$, we do expect similar results though
the peak heights are lower in the former case. See the susceptibility
plots presented later. Our suspicions are that we did not gather
enough data points near $T_c$, leading to less accurate estimates of
the ratio.

Though our numerical results indicates non-Ising behaviour, there may
still be problems.  The phenomena of critical slowing down of the
system dynamics which becomes more significant as we probe closer to
$T_c$ may affect our numerical results.

Unfortunately, we cannot quantify how this phenomena will affect our
results of $\chi$ and $T_c$ near criticality.  As this conflicts with
our need to get a better estimate of the susceptibility peak, we
attempt to counteract via longer running times up to 1 million MCS,
while only up to 500,000 MCS would be more than sufficient to plot the
phase diagrams. This is due to the divergence of the correlation time
near $T_c$, where very long running times would be needed as we go
``closer" to the critical region than could be realised in practice
for large systems.  We are confident that $1 \times 10^6$ MCS used
should be sufficient for $2 \times 32 \times 32$ and $2 \times 64
\times 64$ systems but may not be so for the $2 \times 128 \times 128$
system.  As the algorithm already has almost linear running time, it
would not be trivial to improve upon.  Hence, this huge demand on
computer resources also limits the number of data points we can
collect.

The observed peaks are increasing at a rate higher than (Ising)
expected as $L$ increases and we do not see any reasonable way to
``bring down" the peak heights. For temperatures near criticality the
appropriate entry of the power spectrum we are monitoring (whose time
average is the order parameter) is quite constant but with sudden
drops to zero, like ``pot-holes" in the ground. Note that the drops
as depicted are not as sudden, since we sample the data only every 200
MCS. In our case the entry is for the FE phase. The susceptibility is
known to diverge near criticality, which implies huge fluctuations of
the dominant power spectrum entry. For low (and high) $T$'s, the
constantly high (and low) values gives very small fluctuations and
hence susceptibilities close to zero, as expected and indeed
observed. The above is the general expectations for Ising systems,
where the FE phase's dominance near $T_c$ changes intermittently and
all other ordered phases are negligible.

For our situation, the story is slightly different.  When drops occur
for the FE representation, the entry for AFS (stripped
antiferromagnetic layers) rises.  They are in a way antagonistic to
each other! This curious observation of the possibility that the
dominant phase may occasionally lose out to its local minimum
``sibling" during its evolution towards the steady state speaks of a
non-Ising behaviour. This is only seen near criticality and its power
spectrum entry either stays near its peak value (low $T$) or close to
zero (at very high $T$) elsewhere.

Closer scrutiny of the fluctuation plot (Fig.\ref{fig:fluctuate})
actually reveals that the $L$=128 system near $T_c$ has equilibrated,
since there is no observable time asymmetry.  In fact, the explanation
for the observed $\gamma/\nu$ being more close to the upper limit of
2.0 could be in the plot itself!  This is because we can interpret the
switching of the dominant phase between FE and AFS as a signature of a
first order transition, where $\gamma/\nu$ is exactly 2.0. Hence, the
configuration of the negatively coupled bilayer system could be FE at
moderately low temperatures and as $T_c$ is approached, the AFS phase
becomes significant and competes with FE in the second order {\it
structure-disorder} transition!  This is possible since the AFS phase
is only slightly higher in energy compared with FE and is in fact a
local minimum while FE is the global one. As $T$ is increased further,
the amplitudes of both components were observed to become comparable
till they both become close to zero as for other phases at very high
$T$.

We would like to comment that the driving field does not influence
particle hops across layers.  The disappearance of a particle
poor-rich segregation (across layers) phase could be explained as
chance events, which are frequent due to the high thermal disordering
effects. Particles from the particle rich layer hop to the particle
poor one.  This would bring down the neighbouring particles due to
attractive interactions between particles on the same layer, possibly
resulting in an avalanche.  Without any drive, these clumps of
particles in a generally particle poor region would not have any long
ranged order. However, under the drive, linear interfaces would tend
to result due to particle alignment with the external field. Thus
effect is especially important near $T_c$ where the correlation length
diverges. Locally the rule of having particle-hole pairs across layers
were satisfied by both FE and AFS phases. What resulted was much like
switching between weak forms of the FE and AFS phases, a first-order
transition-like behaviour. Such ability of the lattice gas to switch
between two phases of very close energy does not have a counterpart in
the equilibrium model.

\begin{figure}
\resizebox{8cm}{!}{\includegraphics{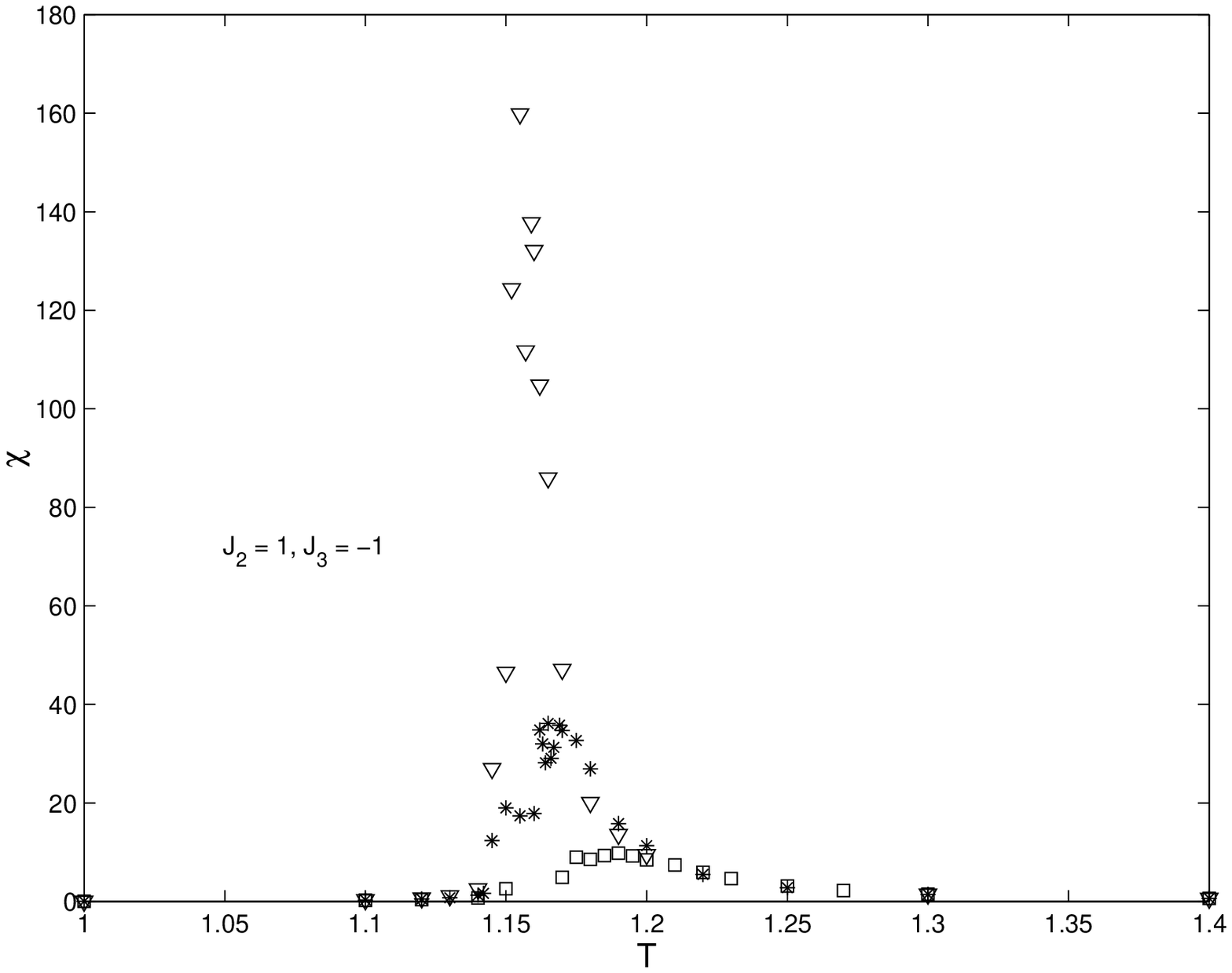}}%
\resizebox{8cm}{!}{\includegraphics{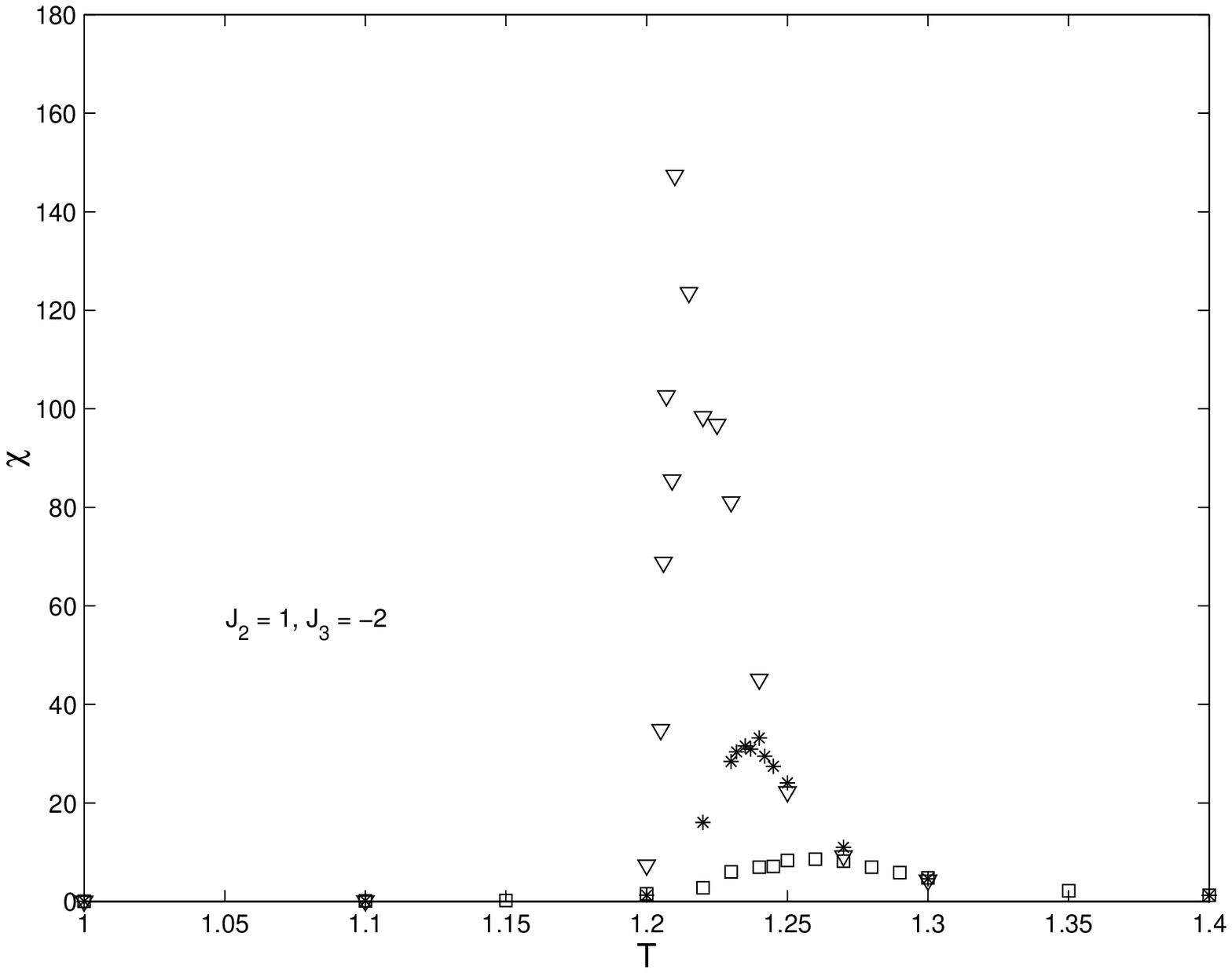}}
\resizebox{8cm}{!}{\includegraphics{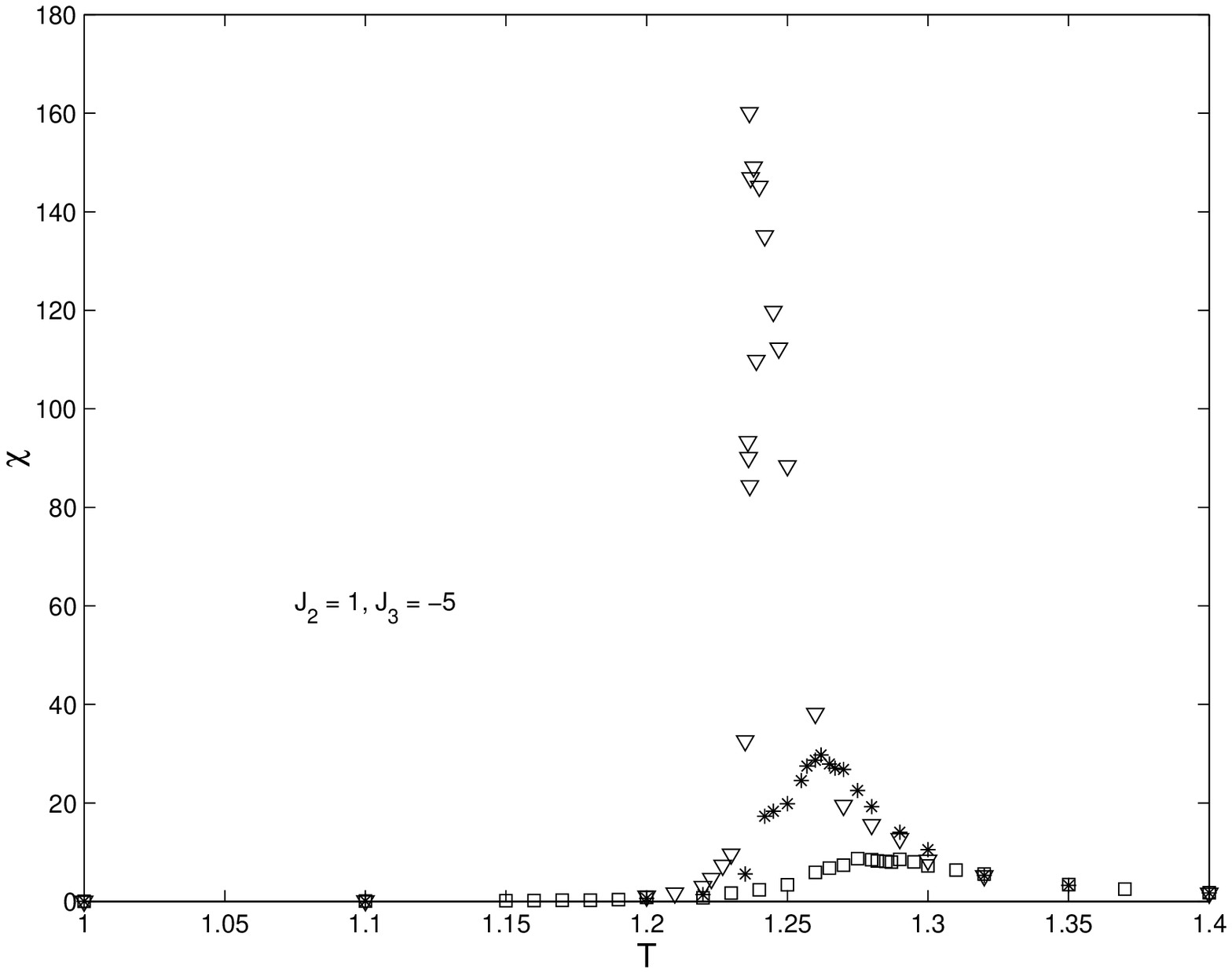}}%
\resizebox{8cm}{!}{\includegraphics{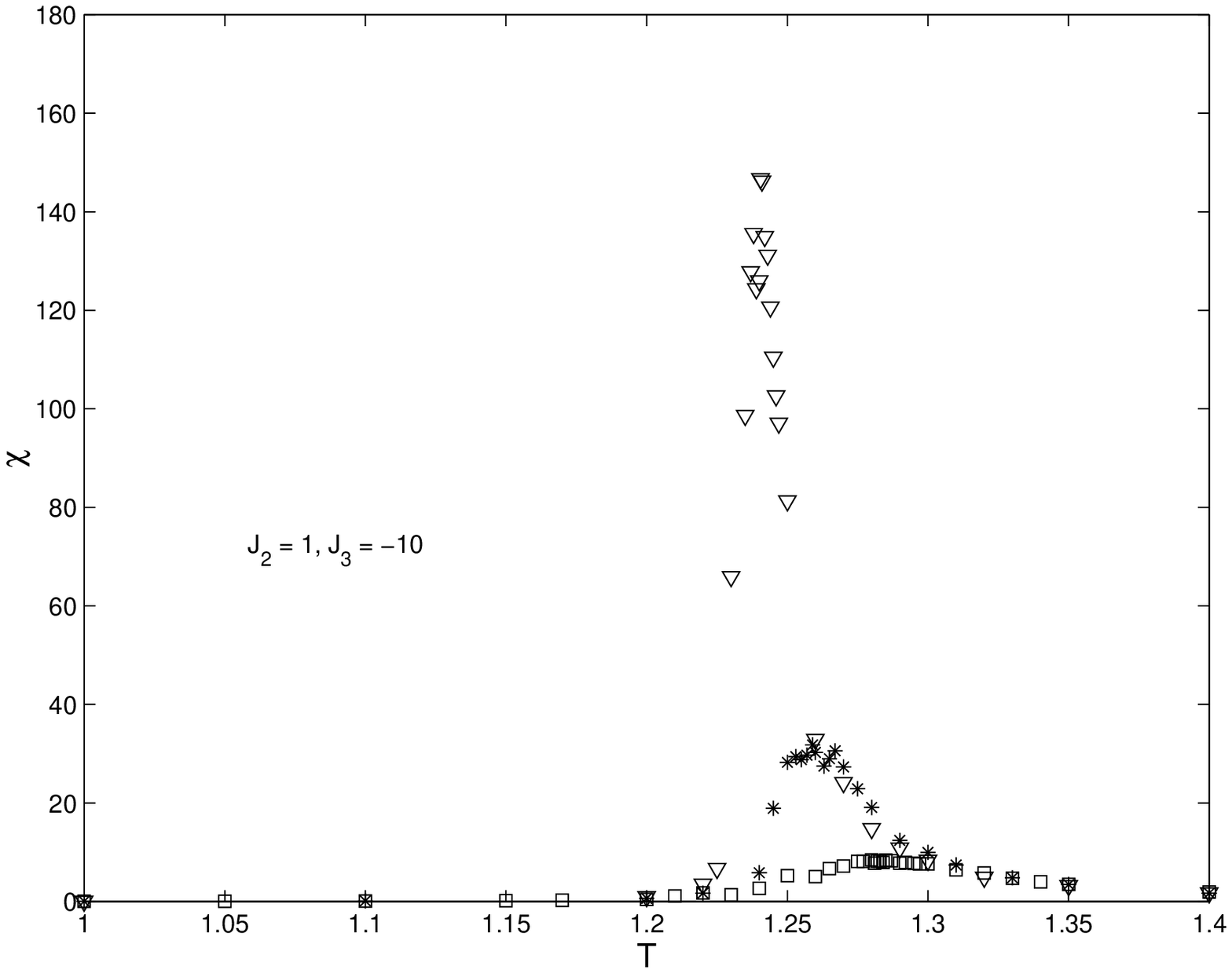}}
\caption{Combined susceptibility plots for driven system with $J_2 =
1$. Note the shift of $T_{peak}$ as $J_3$ becomes more negative, as
shown in the phase diagrams.}
\label{fig:manychisj21}
\end{figure}

\begin{figure}
\resizebox{8cm}{!}{\includegraphics{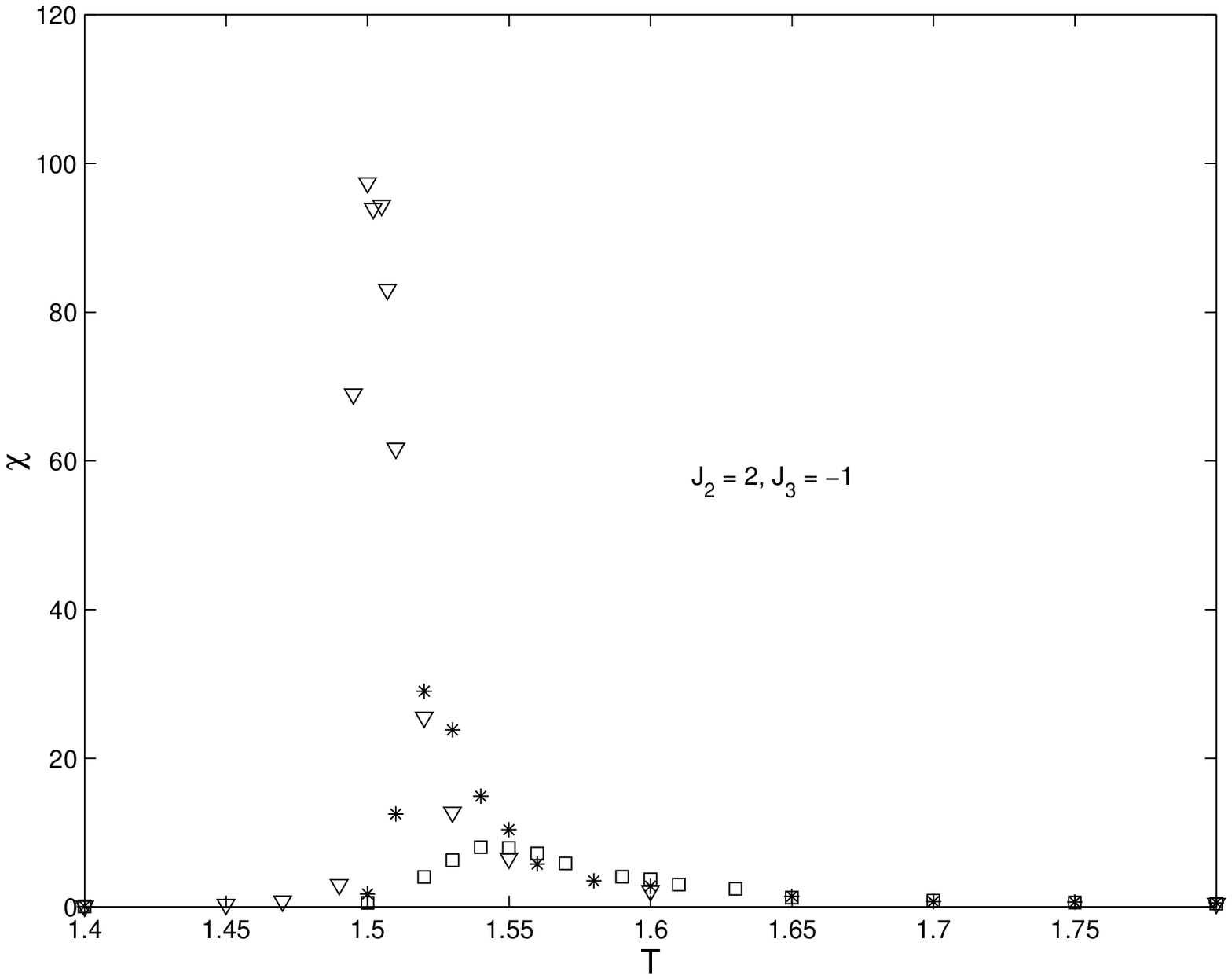}}
\resizebox{8cm}{!}{\includegraphics{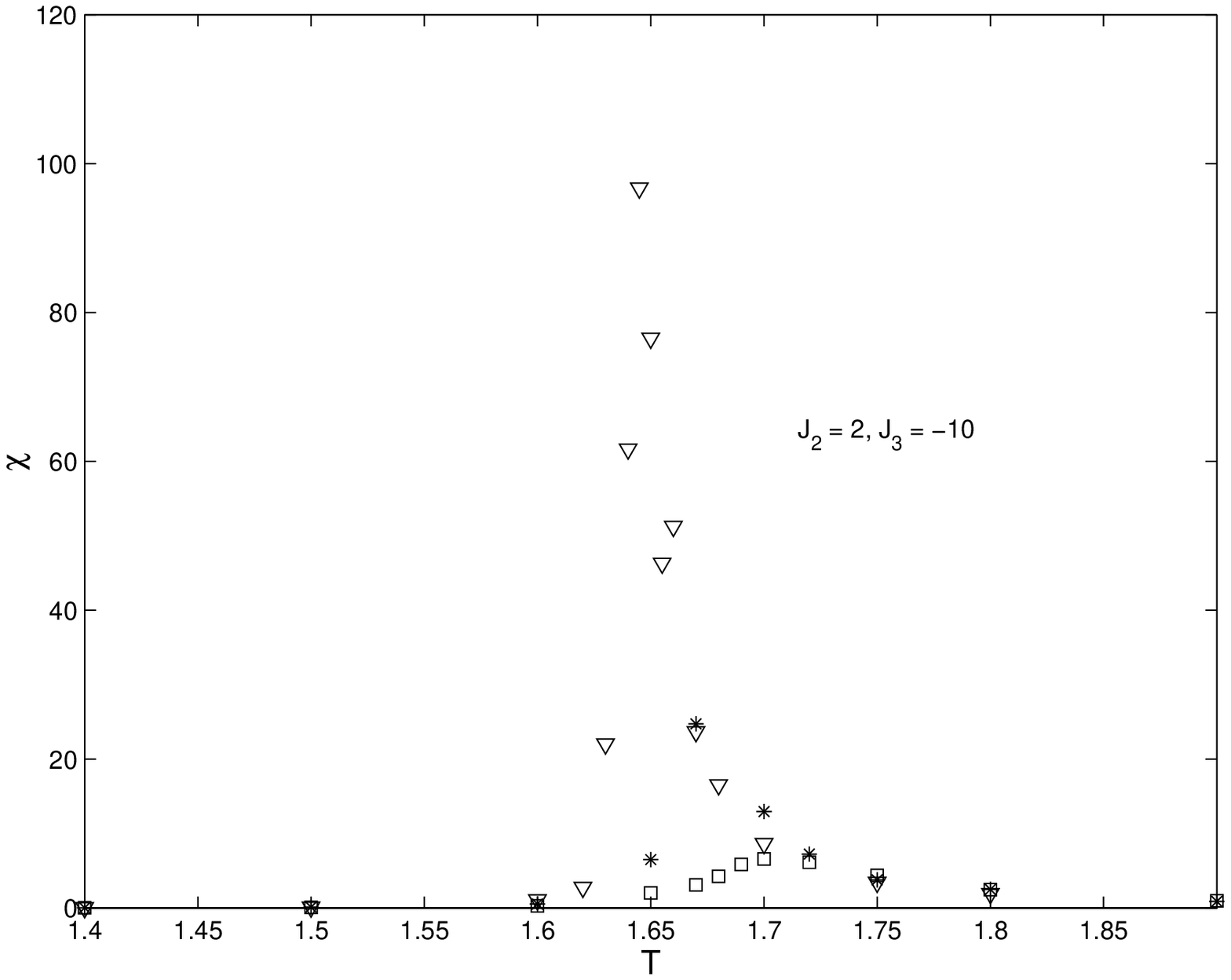}}
\caption{Combined susceptibility plots for driven system with $J_2 = 2$.}
\label{fig:manychisj22}
\end{figure}

Plotting the susceptibility curves for each set of parameter settings
over the different system sizes, we found plots characteristic of
second order phase transitions. See Figures \ref{fig:manychisj21} and
\ref{fig:manychisj22}.  The structure factor data $S(0,0,1)$ (no shown)
at the transitions are smooth and no discontinuities.
This is expected since we monitored the change
of the FE phase as $T$ increases. We would not be able to get any
delta functions characteristic of first order transitions as the
transition between a phase more FE and one more AFS occur during a
single run. What we are implying is that D-FE is second order but near
$T_c$, any disordering effect on the FE structure by the moderately
high temperature is ordered into a AFS-like phase by the large drive
in its direction. This does not occur for undriven systems.  The AFS
phase is not an equilibrium phase by energy arguments since the FE
phase is the more stable one given the same set of conditions, thus we
will not have normal transitions between their fully ordered
forms. The key ingredient is the large, finite driving field which
leads to this nonequilibrium phenomenon.

\section{CONCLUSIONS}

We have attempted to extend the phase diagrams of the bilayer driven
lattice gas for unequal intra-layer attractive couplings. This is in
continuation to the work done by Hill {\it et.al}
\cite{hill:bilayer}. The main findings are that the phase region
occupied by the configuration which consist of ferromagnetic bands
across the layers (S phase) increases in the expense of the other
phase, which is the FE (Filled-Empty) phase. We speculate that the
preference of the S phase over the FE phase by the driving field
increases as the intra-layer coupling traverse to the drive increases.

We also tried to determine the universality class of our bilayer
lattice model with repulsive inter-layer interactions. Starting with
an Ising hypothesis, we found discrepancies of the ratio $\gamma/\nu$
with the Ising value of 1.75. The ratio determined from the peaks of
susceptibility plots according to the finite-size scaling theory is
found to be closer to 2.0. Due to the similarity of the plots with
Ising ones, we assumed $\gamma$ to take the Ising value of 1.75 and
self-consistent plots using the $\gamma/\nu$ ratio independently
determined could be obtained.

The reason for the experimentally determined ratios of $\gamma/\nu$ to
be close to 2.0 is speculated to be due to a first-order transition
like competition of the AFS phase with the FE phase near criticality.
The general D-FE transition should still be second order. This leads
to a non-Ising conclusion.  In fact, this could also explain why the
plots of $T_{peak}$ against $L^{-1/\nu}$ is not linear.

On the other hand, another explanation could be that the scaling is
anisotropic, requiring two correlation length exponents $\nu_{\perp}$
and $\nu_{\parallel}$, associated with the directions perpendicular
and parallel to the driving field, respectively. This could also
explain the nonlinearity of the $T_{peak}$ plots.  However, as is well
acknowledged in the field, this proposal would be very difficult to
investigate.

There is in fact some work on the universality class of bilayered
systems by Marro {\it et.al} in \cite{marro:decoupled2}.  There they
looked at the differences between single and twin-layered driven
lattice gases, where they concluded that the S-FE transition is Ising
in nature. However, no work is done for the D-FE transitions.

\begin{figure}
\resizebox{10cm}{!}{\includegraphics{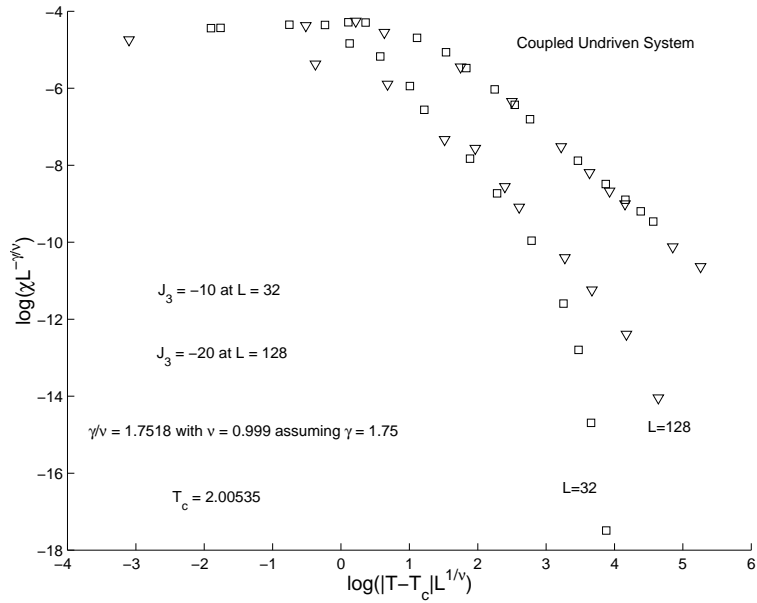}}
\caption{Undriven, coupled case using experimental data for
$\gamma/\nu$ and assuming $\gamma$ = 1.75.}
\label{fig:undriven_coupled}
\end{figure}

On hindsight, we should do a comprehensive study of the undriven case
and compare the current results with it in order to isolate the
effects of the drive. But we expected the bilayered, undriven case to
be well studied and only looked at the case of large interlayer
repulsion, namely the case of $J_3$ = -10 for $L$ = 32 and $J_3$ = -20
for $L$ = 128.  Combining the two sets of data, which is allowed as
the system behaviour should be similar for such large repulsions, we
obtained a $\gamma/\nu$ of 1.7518, which is very close to 1.750 for
Ising. With $\nu$ = 1, $T_c$ of 2.0053 is obtained, which is the
expected result since as $J_3 \rightarrow \pm \infty$, the bilayer
structure becomes irrelevant and the system reduces to a 2-D Ising
system with twice the coupling. This can be understand as cross layer
particle-particle pairs or particle-hole pairs moving in unison in the
2-D lattice.  However, when we attempt to do a ``data-collapse" plot,
the collapse is reasonable but the slope of the top branch is only
1.60! Hence we have a slight consistency problem. See
Fig. \ref{fig:undriven_coupled}.  We can see that the two branches are
not quite parallel, with the lower one giving a slope closer to 1.75.
Further, as the susceptibility plot for $L$ = 128 is not very refined
near the peak, this could introduce errors in the peak
estimation. Also, the run lengths used were only 500,000 MCS in view
of the fact that larger repulsion should lead to faster
equilibration. Nonetheless, the evidence speaks strongly of Ising in
this case.

It is worthwhile to note that though the Ising universality class is
broad, there are exceptions as found in \cite{phil:CML} where $\nu$
could be only 0.89 and in \cite{ferr:ising_fluid} where $\nu$ is 1.35
but in both cases $\gamma/\nu$ is still 1.75. In both cases, the
system has only a single layer and no driving fields are
present. Here, we have a new situation where $\gamma/\nu$ is non-Ising
but $\gamma$ could remain Ising!

Thus, the universality class of the replusive inter-layer bilayer
lattice gas does not belong to the Ising class due to the presence of
two dominant phases near criticality in the approach of the system
towards disorder.  The theoretical and physical ramifications as well
as an analytical understanding are yet to be worked out.

% the END.

% the bibliography
\include{references}

\end{document}

%% file: references.tex
% References for paper by Choon Peng, Chng and Jian-Sheng, Wang.
% Updated: 19/10/1999.
% Physical Review E style: Only books are italized, not journal names.